\documentclass[onecolumn]{article}

\newcommand{\mathsym}[1]{{}}
\newcommand{\unicode}[1]{{}}

\usepackage{amsmath}
\usepackage{tensor}
\usepackage{amssymb}
\usepackage{graphicx}
\usepackage{mathtools}
\usepackage[nopar]{lipsum}

\newcommand{\be}{\begin{equation}}
\newcommand{\ee}{\end{equation}}

\newcommand{\mincir}{\raise
-3.truept\hbox{\rlap{\hbox{$\sim$}}\raise4.truept\hbox{$<$}\ }}
\newcommand{\magcir}{\raise
-3.truept\hbox{\rlap{\hbox{$\sim$}}\raise4.truept\hbox{$>$}\ }}

\def\bea{\begin{eqnarray}}
\def\eea{\end{eqnarray}}

\def\be{\begin{equation}}
\def\ee{\end{equation}}

\PassOptionsToPackage{linktocpage}{hyperref}
\def\case#1/#2{\textstyle\frac{#1}{#2}}

\def\be{\begin{equation}}
\def\ee{\end{equation}}
\def\bea{\begin{eqnarray}}
\def\eea{\end{eqnarray}}

\begin{document}

\title{Persistence in black hole lattice cosmological models}

\author{A. A. Coley:\\
	Department of Mathematics and Statistics, 
	Dalhousie University,\\
	Halifax, Nova Scotia, B3H 4R2, Canada: aac@mathstat.dal.ca}

\maketitle

\begin{abstract}

Dynamical solutions for an evolving multiple network of black holes near a cosmological bounce dominated by a scalar field are investigated.
In particular, we consider the class of
black hole lattice models in a hyperspherical cosmology, 
and we focus on the special case of
eight regularly-spaced black holes with equal masses when the
model parameter $\kappa > 1$. 
 We first derive exact time evolving solutions 
of instantaneously-static models, by utilizing
perturbative solutions of the constraint equations
that can then be used to develop exact 4D dynamical solutions of the 
Einstein field equations. 
We use the notion of a geometric horizon, which can be characterized by curvature invariants, to determine the black hole horizon. 
We explicitly compute the invariants for the exact dynamical models obtained.
As an application, we discuss whether black holes can persist in such a universe that collapses and then subsequently bounces into a new  expansionary phase.
We find evidence that  in the physical models
under investigation (and particularly for $\kappa > 1$)
the individual  black holes do not merge before nor at 
the bounce, so
that consequently black holes can indeed persist through the bounce.

\end{abstract}

%%\eads{\mailto{aac@mathstat.dal.ca}}

%\pacs{98.80.Jk, 04.20.Jb}

\newpage

\section{Introduction}

In bouncing cosmological models, the present expansion of the Universe is assumed to have followed a preceding collapsing phase. The cosmological bounce which bridges these two stages might be caused by either classical or quantum effects \cite{brand}. 
The present expansion phase of the Universe might also eventually recollapse to a  ``big crunch". A cyclic Universe might result if
this is subsequently followed by a bounce.  

In such a bounce scenario,  it is of interest to ask what happens to any population of black holes present. In our Milky Way galaxy, there are billions of stellar-mass black holes, and it is believed that there is also a multitude
of  supermassive black holes at the centers of other galaxies. It is also conceivable that there are ``primordial'' black holes, which formed in the early Universe, which might potentially contribute to any dark matter present \cite{cks}.

In a complete cosmological collapse, it is 
expected that all black holes will merge once the Universe becomes sufficiently compressed. This would also occur in the cosmological bounce model if the matter density at the bounce is high enough. In this merging, if the black holes are distributed randomly with a range of possible masses,
it  might be expected that
merging would occur in a hierarchical manner, in which increasingly larger horizons form around multiple black holes, whereby the characteristic mass of the black holes increases. 
In time the filling factor, $F$, of the black holes (which represents their ratio of size to spacing) will likely reach unity, with the horizons of the individual black holes disappearing completely. Alternatively, in the mathematical idealisation of an exactly regular lattice distribution of black holes (with identical masses), it might be expected that this merger would occur instantaneously at a particular epoch without any preceding hierarchical merging. In either case, there would be a transition in 
which in some appropriate sense the whole Universe turns into a black hole.

In Carr and Coley \cite{cc} (henceforward referred to here as $CC$) the merging of a population of black holes at a cosmological bounce  was discussed. In  $CC$ it was
assumed that all of the black holes have the same mass $M$ and the volume filling factor $F$ is of order $(R_S/L)^3$, where $R_S= 2GM/c^2$ is the Schwarzschild ``radius'' 
or, rather, the Schwarzschild characteristic ``scale'' (where  we will generally set $G=c=1$ hereafter and $L$ is the separation of the black holes). 
Indeed,  $CC$ determined when the value of $F$ for black holes formed in the previous phase (referred to as `pre-crunch black holes' (PCBHs))
would continue to be less than unity at the cosmological bounce, thereby guaranteeing their persistence into the subsequent phase of expansion. 
$CC$  determined the region in which 
the PCBHs that presently contribute to the dark matter survive, 
and computed the minimum possible mass of any black holes formed during the actual cosmological bounce itself (referred to as `big crunch black holes' (BCBHs)). Although similar to primordial black holes (PBHs), these black holes would form immediately prior to rather than just after the big bounce/bang.

$CC$ assumed that the Universe bounces at a density $\rho_B$, which may be of order the Planck density
but, in principle, $\rho_B$ could be much less. Now, a
spherical sector of mass $M$ forms a black hole 
when it descends within its Schwarzschild radius, with density  
$\rho_{BH}  \sim 10^{18} \left({M}/{M_{\odot}}\right)^{-2} \mathrm{g \, cm}^{-3}$.  
A BCBH (which forms during the bounce) has $\rho_{BH}$ which is 
necessarily bigger than the cosmological density at the bounce, and we
obtain a {lower} limit on the black hole mass 
$M   \sim \left({\rho_P}/{\rho_B} \right)^{1/2} M_P$,
where $M_P$ is the Planck mass.
This limit also applies to when pre-existing PCBHs  (which form before the bounce) 
might lose their own identity in the merger with other PCBHs.  
If $F_B$ represents the fraction of the Universe's density  in PCBHs at the instant of the bounce, then the average distance between the black holes at the bounce, $R_{\mathrm{sep}}$,
is less than the size of each individual black hole; 
$R_{\mathrm{sep}} \lesssim  2GM/c^2$. The merging condition 
$M \gtrsim  F_B^{-1/2} M_{\mathrm{min}}$
can then be interpreted as a minimum on the fraction of the Universe contained in the black holes. 
Therefore, a range of black hole masses in which BCBHs can form but PCBHs do not merge is obtained.
The fraction $F_B$ is constant in a matter-dominated epoch, but will decrease during collapse in a radiation-dominated epoch; however, 
the fraction of the material content of the Universe in black holes can still be computed $CC$.

There is a variety of constraints on the quantity and mass of non-evaporating black holes (i.e., those with mass larger than $M \sim 10^{15}$g) from
lensing, dynamics and astrophysics
~\cite{cks}.  Constraints on evaporating black holes
are not relevant for PCBHs, but they might be for BCBHs. In
particular, an important dynamical constraint is obtained from large-scale structure formation due to the fluctuations in the black hole number density \cite{Meszaros}.

The analysis of $CC$ may be sufficient in some situations.
However,  it cannot be reliable in general since each  black hole does not have the Schwarzschild radius when the black holes get sufficiently close together (and it is
certainly not valid as $F \rightarrow1$). 
Indeed, the very
definition of  a black hole in a cosmological background is questionable. 
Since the notion of an event horizon cannot be used (e.g., there may be no spatial infinity), often the behavior of the so-called apparent horizon \cite{AshtekarKrishnan}
is studied. But, in the cosmological context, the apparent horizon of a black hole
can be different from the Schwarzschild value.  As a simple example, if the background universe has a non-vanishing cosmological constant, $\Lambda$, the apparent horizon will depend on both $M$ and $\Lambda$ and is smaller than $R_S$, even when assuming spherical symmetry.  In addition, the 
assumption of spherical symmetry in the case of the apparent horizon
is violated when
the filling factor approaches $1$. Indeed, in the idealisation that the black holes 
are  configured in a perfect lattice, they tend to ``cubes'' rather than ``spheres''.

The considerations of 
$CC$  were essentially heuristic and not based on any rigorous computations. In a subsequent paper \cite{bounce} (henceforward referred to as $CCC$ here),
utilizing earlier work of Clifton {\it et al.}~\cite{clifton2},
exact solutions describing a regular black hole lattice in a cosmological background dominated by a dynamical scalar field at the bounce were derived.
In particular, the question of  whether black holes can persist in a Universe that collapses and then subsequently bounces into a new  phase of expansion
was studied 
by investigating the maximal number of black holes that can keep their individual identity through the bounce \cite{bounce}.

\newpage

\subsection{Black hole lattice models}

A set of models of interest are the black hole lattice (BHL) models \cite{bruneton,num0} in which the matter content of the universe is 
divided into discrete cells, each of which is then  modeled as a black hole. 
The resulting spacetimes are explicitly highly inhomogeneous on small scales,
but they  are spatially homogeneous and isotropic on large scales \cite{DurkClif}. 
These spacetime models can be used to characterize
the coarse-graining of spacetime  \cite{rvdh}, and do not necessarily behave dynamically in precisely the same way as the {\it exactly} spatially homogeneous and isotropic
Friedmann-Lema\^{i}tre-Robertson-Walker (FLRW) solutions.

A property of these particular solutions is that the number of black holes ($N$),
which are often assumed to be  spaced regularly, is finite and reasonably small. For example, in a closed model, $N$ has the possible values $5, 8, 16, 24, 120, 640$.  
(However, $N$ could be larger if the assumption of uniformity is dropped and could be arbitrarily large in a flat universe).
The associated cell size to model the Universe at present is consequently  $N^{-1/3}R_{PH} \gtrsim  600$~Mpc.
This, of course, does not model the actual situation in our Universe, in which the total number of supermassive black holes is perhaps of the order $10^{10}$ (e.g., one per galaxy) and  stellar mass black holes might  number on the order $10^{20}$.

Recently, the effects that mass-clustering has on the scale and characteristics of  cosmology within the class of BHL models 
was studied \cite{DurkClif}.
Although these models are too simplistic to model the complex structures in our actual Universe, they do equip us with a means to investigate some aspects of coarse-graining and the averaged expansion \cite{rvdh}.
The BHL models were generalized
to include clusters of masses, and consequently allow for the dynamical effects of cosmological structures to be investigated in a well-defined and non-perturbative manner. The  cosmological implications of the formation 
of structure on the total amount of mass in each of the clusters and of the energy of the interactions within and between clusters
was studied. 
The positions of the shared horizons that surround groups of sufficiently close black holes were 
investigated.
It was found that a common apparent horizon around the clustered black holes would 
appear when the black holes  become sufficiently close together (in addition to the apparent horizon of each 
individual  black hole).

\subsubsection{CCC BHL bounce model}

$CCC$ \cite{bounce} obtained a class of  exact ``time-evolving'' solutions to Einstein's constraint equations modeling the spacetime geometry at the surface of minimum of cosmological expansion. 
These models contain a regular BHL and a scalar field at the maximal collapse.
The associated initial data can then be utilized to uniquely determine the  evolution in (both) the expansion (and collapse) 
{directions}, and particularly the dynamical 
evolution of the hyperspherical cosmological models;
consequently they constitute a well-defined class of cosmological models. 
The actual initial data can itself be used to determine the locations of the  black hole apparent horizons and utilized to compute the distance between neighbouring black holes, which then determines the fraction of space filled by black holes at the bounce in terms of the optimal energy density attained there.

These models clearly illustrate the fact that
it is possible for black holes to persist through a cosmological bounce.
In particular, $CCC$ considered hyperspherical cosmological models.  
It was determined that the model parameter, $\kappa$, is constrained to lie within a particular range of values if both the black holes have positive mass and also that the value of the scalar field is sufficiently large to dynamically ensure a bounce.   

Black holes in scalar field cosmologies with spatial flatness (and solutions with $V \neq 0$), and 
the bounce-merger conditions for persistent black holes through a cosmological bounce in higher-dimensional spaces,  
were also considered \cite{bounce}.
Apparently, there are 
difficulties producing physical  bouncing cosmological models containing non-merging black holes of the type discussed in $CCC$ in spatial dimensions greater than five.
There continues to be  interest in seeking  time-evolving solutions in models in which a multiple number of distinct black holes persist through a non-time-symmetric bounce.

Returning to the question of the possible maximal {fraction} of space filled by PCBHs or the number of BCBHs that can form as the minimum of expansion is approached,
$CCC$ were able to accurately compute $F$ for the BHL  cosmological models. Indeed, it was found that there exist solutions in which the universe bounces before $F$ reaches unity.  Therefore, $CCC$ obtained the important result that {\it there do exist exact solutions in which black holes
persist through a bounce.} In addition, the expression found for $F$ at the bounce is not too dis-similar from the heuristic result found in the $CC$ analysis for $F \ll 1$.
Although the $CCC$ models are  not formally valid for large values of $\kappa$, 
there is an indication from $CCC$ that as $\kappa$ increases  $F $ approaches but never exceeds unity at the bounce. As $F \rightarrow 1$, the majority of the universe is in the black holes rather than ``in the Friedmann region''. This is not dis-similar to the case of a black hole in a positive curvature Friedmann geometry, in which $M$ approaches zero as the size of the black hole tends to that of the background. Therefore, it is not implausible that the filling factor $F$ can never reach $1$.  

One of the primary motivations of the current work is to investigate whether something may prevent the filling factor ever reaching $1$ in the bouncing models under investigation. Indeed,
perhaps the size of the individual  black hole becomes smaller than the standard Schwarzschild value or the effective black hole mass approaches zero. 
In \cite{bounce} the area method was used to compute the apparent horizon.  But this method is not reliable, particularly for larger values of $\kappa$. Therefore, we shall
utilize the notion of a geometric horizon here.

\newpage

\subsection{Geometric horizon}

The event horizon of a black hole spacetime in General Relativity
(GR) is defined as the boundary of the non-empty complement of the
causal past of future null infinity (that is, the sector for which signals originating in the
interior will never escape).  Note that the global
behaviour of the spacetime must be known in order for the event horizon to be determined locally \cite{AshtekarKrishnan}.
However, to examine the interaction of realistic black holes with their environment, 
in which the  black holes are dynamical, a local characterization that 
does not necessarily depend on the event horizon is needed.
As a result, when considering time-dependent situations
rather than
stationary black holes, an event horizon is often replaced in practical applications by an apparent horizon, which is defined as the locus of the vanishing expansion of a null geodesic
congruence starting within a trapped surface S (with a spherical topology) \cite{RPenrose}.
Indeed, in numerical studies of collapse,
the apparent horizon is a more
practical surface to track  \cite{JThornburg}. A
related concept is marginally outer trapped surfaces (MOTS),
which are defined to be two-dimensional (2D) surfaces for which
the expansion of the outgoing null vector normal to the
surfaces is zero. Assuming a smooth dynamical evolution, these 
2D surfaces can be combined to construct a 3D surface
known as a marginally trapped tube \cite{BoothFairhurst}. 
Unlike the event horizon, the apparent horizon is quasi-local. But it is
intrinsically foliation-dependent, which can lead to ambiguities 
and implementation difficulties. 
As a result, it is of great import to define other surfaces that are characterized in an
invariant manner.

%%%%%%%%

In the case of a stationary black hole spacetime, knowledge of the Killing vector field
acting as the null generator on the event horizon then allows the horizon to be characterized
locally. This implies that there are consequently
scalar polynomial (curvature) invariants (SPIs) which are zero
on the stationary horizon  \cite{PLB}.
In particular, a special
set of SPIs vanish on the horizon since the curvature tensor and its covariant derivatives must be
of type {\bf II/D} relative to the alignment classification \cite{CMPP} there.
It has subsequently been conjectured that a dynamical black hole admits a quasi-local hyper-surface,  
called a geometric horizon (GH), on which the curvature tensor (and its first covariant derivative) are
algebraically special \cite{PLB}.
Indeed, there are examples of time-evolving black hole spacetimes in which a
GH is defined (for example, in the case of
spherical symmetry).

That is, we shall utilize the concept of a GH, which can
be characterized invariantly  by the vanishing of a defining set of scalar curvature invariants \cite{PLB}, 
as an alternative method to using an apparent horizon. In particular, we shall use the GH to characterize the horizon in the BHL models under study here.

\newpage

\subsection{Overview}

We wish to study whether black holes can persist in a  collapsing universe that subsequently bounces into a new expansionary phase. 
In particular, we are interested in the black hole
number density  as the minimum of expansion is approached and whether it is
possible that 
the filling factor may be prevented from ever reaching unity. For example, we wish to pursue the 
possibility that the size of each individual  black hole is smaller than the usual Schwarzschild value.

Therefore, we seek an  alternative method for determining the black hole horizon,
and we shall utilize the notion of a GH, which can be simply determined by curvature invariants.  This necessitates
looking for time-evolving multiple black hole solutions. 
We shall follow
\cite{bounce} and consider a scalar field cosmology to model the matter content of such a universe close to the bounce, and seek solutions representing a dynamical network of black holes.

While the phantom
scalar field bounce model considered here (in which the scalar potential is set to zero -- see later) is not perhaps the most physically realistic
matter model,  it does provide exact bounce solutions in which persistence occurs and
which can be primarily studied analytically (hence  providing some basic underlying
understanding). More realistic bounce models can be considered,
based on a more effective field theory (which, e.g., does not suffer from a scalar field instability problem due to a violation of the energy conditions), including models which include a cosmological 
constant or a non-minimally coupled scalar field \cite{Bronnikov}. Such models will still lead a well-posed initial value problem (which can be defined in an analogous manner to that
done here) that will lead to bouncing cosmologies with similar persistence properties,
but these models can only be studied numerically.

In more detail,
we shall begin by presenting and reviewing the class of
black hole lattice models in a hyperspherical cosmology. 
We derive exact time evolving solutions 
of instantaneously-static models, by employing
perturbative solutions of the constraint equation
that can then be utilized to develop exact 4-dimensional (4D) dynamical solutions of the Einstein field equations (EFEs). Although
these solutions are very idealistic, they do, however, model a possible realistic 
scenario in which a number of distinct black holes persist through a cosmological 
bounce. 

We focus on the particular case of
eight regularly-spaced black holes with equal masses, and we consider the relevant case of when the model parameter (see section 3) $\kappa > 1$. 
We then compute the invariants for these exact solutions  necessary to determine the conditions for a GH explicitly. We conclude with  a summary and a  
discussion of the results.

\newpage

\section{BHL models: Constraint and evolution Eqs.} 

We will display the EFEs for a radiation fluid and a scalar field, appropriate for investigating the spacetime geometry in the vicinity of a bounce. 
We will first present the constraint and evolution equations in a general form and discuss how a bouncing cosmological model with black holes can be determined as an initial value problem, where the appropriate initial data is given at the bounce.

Similarly to normal early-universe FLRW cosmology, the scalar field is anticipated to dynamically dominate at early times but to subsequently become negligible in comparison to the energy-momentum in radiation and dust at later times. 
This implies that in the case of a regular lattice of black holes and a scalar field is considered, the 
subsequent late-time dynamics of the spacetime should approach that of one simply consisting of an array of black holes. This is pertinent because  there is a time-symmetric initial value problem for a BHL 
in the case of a scalar field with a negative coupling constant in which
the black holes 
move relative to each other as the space evolves away from the initial time-symmetric surface. Indeed, due to the time symmetry on the initial surface, the constraint equations can be formulated in a linear form, thereby allowing for the initial data for an arbitrary number of black holes to be constructed by addition.

This initial data problem was studied in  \cite{clifton2,DurkClif}. In general, the models can be investigated by  separating the governing EFEs into sets of constraint and evolution equations relative to a $1 + 3$ decomposition. The intrinsic geometry and the extrinsic curvature of an initial hyper-surface then constitute sufficient initial data to ensure a unique evolution. Therefore, if the constraint equations can be solved at some initial time, then there is sufficient data for the geometry of the complete spacetime to be determined.

In the next subsection we closely follow the approach of  \cite{bounce} and we shall repeat the relevant equations therein, in order for the current paper to 
be self-contained.

\subsection{Field equations and time-symmetric initial data}

The EFEs for radiation and a scalar field are given by:
\be
\label{fe}
G_{ab} = \mu T^{\varphi}_{ab} \,  + \, T^{\gamma}_{ab} \, ,
\ee
where $\mu$ is a  negative coupling constant (necessarily to produce a bounce) and units are selected 
so that $8 \pi G = c=1$. $T^{\varphi}_{ab}$ is the energy-momentum tensor of the scalar field:
\be
T^{\varphi}_{ab} = \nabla_a \varphi \nabla_b \varphi -\frac{1}{2} g_{ab}   \nabla_c \varphi \nabla^c \varphi - g_{ab}  V \, ,
\ee
where $V \equiv V(\varphi)$ is the self interaction potential of the  scalar  field. $T^{\gamma}_{ab}$ is the energy-momentum tensor of a  separately conserved radiative matter field. The contracted second Bianchi identity then implies the evolution Eq. for the scalar field:
\be
\label{boxvarphi}
\nabla^a \nabla_a \varphi = \frac{dV}{d\varphi} \, ,
\ee
where $\nabla^a \nabla_a$ is the covariant d'Alembertian operator.

Choosing a time-like vector field, $u^a$, we can then decompose $T^{\varphi}_{ab}$ into an effective energy density, $\rho^{\varphi}$, an isotropic pressure, $p^{\varphi}$, and momentum density, $j_a^{\varphi}$:
\bea
\label{rho}
\rho^{\varphi} &\equiv& u^a u^b T^{\varphi}_{ab} = \frac{1}{2} \dot{\varphi}^2 + \frac{1}{2} D^a \varphi D_a \varphi  +V(\varphi)\, \\
\label{p} p^{\varphi} &\equiv& \frac{1}{3} h^{ab} T^{\varphi}_{ab} = \frac{1}{2} \dot{\varphi}^2 - \frac{1}{6} D^a \varphi D_a \varphi -V(\varphi) \,\\  
\label{j}
j_a^{\varphi} &\equiv& - h_a^{\phantom{a} b} u^c T^{\varphi}_{bc} = - \dot{\varphi} \, D_a \varphi \, ,
\eea
where an ``over-dot'' denotes $u^a \nabla_a$ and $h_{ab} \equiv g_{ab} +u_a u_b$
has been introduced, so that $D_a \equiv h_a^{\phantom{a} b} \nabla_b$ denotes the projected covariant derivative orthogonal to $u^a$. In addition, the effective energy density and effective isotropic pressure of a  radiation perfect fluid are given by:
\bea
\rho^{\gamma} &\equiv& u^a u^b T^{\gamma}_{ab} \, , \qquad p^{\gamma} \equiv \frac{1}{3} h^{ab} T^{\gamma}_{ab} \, ,
\eea
where $p^{\gamma} = \frac{1}{3} \rho^{\gamma}$ and $j_a^{\gamma} \equiv - h_a^{\phantom{a} b} u^c T^{\gamma}_{bc} = 0$.

From the embedding Eqs. and the EFEs (\ref{fe}),  the Hamiltonian and momentum constraint Eqs. for the spacetime can be written as:
\bea
\label{hc}
&&\mathcal{R} + K^2 - K_{ab} K^{ab} = 2 u^a u^b G_{ab} = 2 \mu \rho^{\varphi} + 2 \rho^{\gamma}  \,  \\
\label{mc}
&&D_b K^b_{\phantom{b} a} - D_a K = - h^b_{\phantom{b} a} u^c G_{bc} = \mu j_a^{\varphi} \, ,
\eea
where $K_{ab} \equiv - h_a^{\phantom{a} c} h_b^{\phantom{b} d} \nabla_{(c} u_{d)}$ is the extrinsic curvature of the initial hyper-surface, $K \equiv K^a_{\phantom{a} a}$ and 
(assuming an irrotational $u^a$)
$\mathcal{R}$ denotes the Ricci curvature scalar of this intrinsic 3-surface. These Eqs. must necessarily be satisfied on this initial hyper-surface in order for there to be a solution of the EFEs.

We can now decompose the conservation Eqs. for the scalar field and radiation. 
The constraint and evolution Eqs. for  the radiation fluid are, respectively,
\bea
\label{radcon}
D_a \rho^{\gamma} + 4 \rho^{\gamma} \dot{u}_a =0 \, , \qquad
\dot{\rho}^{\gamma} = \frac{4}{3} K \rho^{\gamma}
\eea
where  $u^a$ has been chosen to be comoving with the radiation fluid. Defining
\be
\label{varphivar}
 \psi_a \equiv D_a \varphi \, ,
\ee
the propagation Eq. (\ref{boxvarphi}) for the scalar field  can then be decomposed into evolution and constraint Eqs. relative to $u^a$:
\bea
\label{evo1}
\dot{\Pi} &=& D_a \psi^a+ K \Pi + \dot{u}_a \psi^a - V^{\prime}(\varphi) \, \\
\dot{\psi}_a &=& D_a \Pi + \dot{u}_a \Pi + u_a \dot{u}^b \psi_b +   K^a_{~b}\psi_b   \, \\
\dot{\varphi} &=& \Pi \, ,
\label{evo3}
\eea
where the last Eq. serves to define $\Pi$. Consequently, Eq.~(\ref{varphivar}) is the one remaining constraint Eq., while Eqs.~(\ref{evo1})-(\ref{evo3}) provide the evolution Eqs. for $\Pi$, $\psi_a$ and $\varphi$. We have thus
completed an appropriate initial value  formulation.

We now investigate the time-symmetric case in which the initial data satisfies $K_{ab}=0$. From Eqs.~(\ref{j}) and (\ref{mc}), we immediately obtain
\be
\dot{\varphi} D_a \varphi =0 \, ,
\ee
and hence  either $\dot{\varphi}=0$ or $D_a \varphi=0$ on the initial hyper-surface. If both are satisfied simultaneously, then the  evolution Eqs. for the scalar field imply that $\varphi$ vanishes everywhere, which corresponds to the trivial case in which there is, in fact,  no scalar field present. The case in which $\dot{\varphi}$ vanishes was studied in \cite{bounce}. 
We shall investigate
the remaining case (of most interest for cosmology) in which $D_a \varphi=0$ henceforward in this paper.
If $D_a \varphi =0$, it follows that on the initial hyper-surface $\varphi$ is constant, and from Eqs.~(\ref{rho}) and (\ref{p}) we then obtain $\rho^{\varphi} = \frac{1}{2} \Pi^2 +V$ and $p^{\varphi} = \frac{1}{2} \Pi^2-V$. The time-symmetry on the initial hyper-surface then implies that $\dot{\rho}^{\varphi}= \Pi (\dot{\Pi} +V^{\prime})=0$ and $\dot{p}^{\varphi} = \Pi (\dot{\Pi} -V^{\prime})=0$. Since a non-vanishing $\Pi$ is required in order for a non-vacuum solution to exist, we then have that $\dot{\Pi}=V^{\prime}=0$ (consistent with Eq. (\ref{evo1})). Therefore, in this case we obtain a solution in which  initially $\varphi$ is spatially homogeneous, but $\dot{\varphi}$ is spatially inhomogeneous and $\ddot{\varphi}$ is zero. It then follows that $V$ is constant and of the form of a cosmological constant.

To produce a solution to the Einstein-scalar equations, it 
remains for the  
the Hamiltonian constraint to be satisfied. When $D_a \varphi = V =0$, 
Eq. (\ref{hc}) can be written as:
\be
\label{constraintR}
\mathcal{R} = \mu \Pi^2 + 2 \rho^{\gamma} \, .
\ee
Since $\Pi$ does not appear in the scalar field constraint Eq., it can consequently be freely specified. A solution for the geometry of an initial hyper-surface satisfying Eq.~(\ref{constraintR}) for a given value of $\Pi$  then leads to a complete solution to all of the constraint Eqs. and consequently to the full initial data necessary to determine the entire subsequent evolution.

In order to obtain explicit solutions with constant conformal curvature, we take the line-element for the time-symmetric initial hyper-surface to be:
\be
\label{le}
ds^2 = \Omega^4 d\tilde{s}^2 \, ,
\ee
where $d\tilde{s}^2$ is the metric for a 3-space of constant  curvature. We have
\be
\label{conformal}
\mathcal{R} = \frac{\tilde{\mathcal{R}}}{\Omega^4} - \frac{8}{\Omega^5} \tilde{D}^2 \Omega \, ,
\ee
where $\tilde{\mathcal{R}}$ is the constant Ricci curvature scalar of the conformal space and $\tilde{D}$ is the covariant derivative on it. Combining these Eqs. then yields
\be
\label{constraint}
\tilde{D}^2 \Omega - \frac{\tilde{\mathcal{R}}}{8} \Omega=  -\frac{(\mu \Pi^2 + 2 \rho^{\gamma})}{8}   \Omega^5  \, .
\ee
For a vacuum spacetime, in which $\mu\Pi^2 + 2 \rho^{\gamma}=0$, this Eq. is linear in $\Omega$, and we can seek a multi-black hole solution by superposition \cite{DurkClif}.  Eq.~(\ref{constraint}) can be made
linear in $\varphi$
by an appropriate choice for $\Pi$, which can be then exploited to obtain a solution for black holes in a hyperspherical cosmology. From the EFEs, it is clear that a bounce can only be obtained when the 
scalar field is massless with a negative energy density
(i.e., when $\mu <0$ and $V=0$).

\newpage

\section{Black holes in a hyperspherical cosmology}
\label{sec:sphere}

We  consider a hyperspherical cosmological model, which has a mathematically simpler initial value problem formulation. The conformal line-element in Eq. (\ref{le}) 
($\tilde{h^0}_{\alpha \beta}$) is taken to as:
\be
\label{hs}
d\tilde{s}^2 = dr^2 + \sin^2 r \left( d\theta^2 + \sin^2 \theta \, d \phi^2 \right) \, ,
\ee
where the Ricci scalar is $\tilde{\mathcal{R}}=6$ and the Hamiltonian constraint (\ref{constraint}) can be written:
\be
\tilde{D}^2 \Omega = \left( \frac{3}{4} -\frac{1}{8} (\mu \Pi^2 +2 \rho^{\gamma}) \Omega^4 \right) \Omega   \, .
\label{ham}
\ee
We select the value of $\Pi$ on the  initial hyper-surface as
\be
\label{c}
\Pi^2 = \frac{8}{\mu \Omega^4} \left( \frac{3}{4} - \kappa \right) -\frac{2 \rho^{\gamma}}{\mu} \, ,
\ee
where $\kappa$ is an arbitrary constant (as defined below in
(\ref{helmholtz}), and which constitutes the essential parameter of the BHL and consequently the resulting cosmological model),
so that the constraint Eq. becomes
\be
\label{helmholtz}
\tilde{D}^2 \Omega = \kappa \, \Omega \, ,
\ee
where $\kappa > 3/4$ if both $\mu <0$ and $\rho^{\gamma}=0$. This equation is, in fact, the Helmholtz Eq. on a 3-sphere, for which solutions are known. And since it is linear in $\Omega$, multi-black hole solutions can be constructed by superposition. If $\kappa = 3/4$, then the sum of the energy densities of the scalar field and the radiation  field is zero on the initial hyper-surface.

\paragraph{Solutions for the conformal factor:}

There are solutions to Eq. (\ref{helmholtz}) of the form:
\be
\label{omi}
\Omega_i (r) = \alpha_i \, \frac{\cos (\sqrt{1- \kappa} \, r)}{\sin r} + \gamma_i \, \frac{\sin (\sqrt{1- \kappa} \, r)}{\sin r} \, ,
\ee
where $\alpha_i$ and $\gamma_i$ are constants (and implicitly we assume that $\kappa < 1$ here; the $\kappa > 1$ case is discussed later). 
The location of the source is taken as $r=0$,
where these solutions diverge. The geometry is smooth at all other points if  the first derivative of $\Omega$ is single-valued. Applying this condition at $r=\pi$ implies that $\gamma_i = - \alpha_i \cot ( \sqrt{1- \kappa}\,  \pi)$, and hence
\be
\label{cont111}
\Omega_i(r) = \alpha_i \frac{\sin (\sqrt{1-\kappa}  \, (\pi-r))}{\sin (\sqrt{1-\kappa}  \, \pi) \sin r} \, .
\ee
This Eq. gives the contribution to the conformal factor due to a single point-like source at $r=0$. For $N$ different sources, located at arbitrary positions on the hypersphere, the corresponding solution can be written as
\be
\label{cont222}
\Omega (r, \theta, \phi) = \sum_{i=1}^N \alpha_i \frac{\sin (\sqrt{1-\kappa}  \, (\pi-r_i))}{\sin (\sqrt{1-\kappa}  \, \pi) \sin r_i} \, ,
\ee
where $r_i$ is defined to be the radial coordinate in Eq.~(\ref{hs}) after rotating the hypersphere (and hence the location of the $i$th source is  $r_i=0$).

The bare mass of each of the mass sources is obtained by comparison with terms that occur in a time-symmetric slice in the exact Schwarzschild spacetime \cite{DurkClif}.
The {\em{proper mass}} of each individual black hole is consequently obtained by a rotation of the coordinates so that the targeted black hole is centered at $r=0$, and by a 
comparison with the leading-order terms in the line element to that of the exact Schwarzschild metric in the limit as $r \rightarrow 0$.

The $r_i \rightarrow 0$ limit of Eq.~(\ref{cont222}) is given by:
\be
ds^2 \rightarrow \left( \frac{\alpha_i}{r_i} +A_i \right)^4 \left( dr_i^2 +r_i^2 (d\theta^2 + \sin^2 \theta \, d \phi^2 ) \right) \, ,
\ee
where
\be
A_i \equiv - \frac{\alpha_i \sqrt{1-\kappa}}{ \tan (\sqrt{1-\kappa}  \, \pi)} + \sum_{j\neq i} \alpha_j \frac{\sin (\sqrt{1-\kappa}  \, (\pi-r_{ij}))}{\sin (\sqrt{1-\kappa}  \, \pi) \sin r_{ij}} \, 
\ee
and $r_{ij}$ is the coordinate distance to the $j$th source from the one located at $r_i=0$. 
(In the above we take the limit as $r_{ij} \rightarrow r_i$ for degenerate points; e.g.,
the antipodal point in the 8BH example below.) A new radial coordinate is defined  by $r^{\prime}_i \equiv \alpha_i^2 / r_i$, so that
\be
ds^2 \rightarrow \left(  1 + \frac{4\alpha_i  A_i}{r_i^{\prime}} \right) \left( dr^{\prime 2}_i +r_i^{\prime 2} (d\theta^2 + \sin^2 \theta \, d \phi^2 ) \right) \, .
\ee
When compared with the exact Schwarzschild solution as $r \rightarrow \infty$,
\be
\label{schwarz}
ds^2 \rightarrow \left( 1 + \frac{2 m}{r} \right) \left( dr^2 + r^2 (d\theta^2 + \sin^2 \theta d \phi^2) \right) \, ,
\ee
we then obtain the proper mass of the $i$th source:
\be
m_i = 2 \alpha_i A_i = - \frac{2 \alpha_i^2 \sqrt{1-\kappa}}{ \tan (\sqrt{1-\kappa} \pi)} + 2 \sum_{j\neq i} \alpha_i \alpha_j \frac{\sin (\sqrt{1-\kappa}  \, (\pi-r_{ij}))}{\sin (\sqrt{1-\kappa}  \, \pi) \sin r_{ij}} \, .
\ee
We remark that the mass of each individual source depends on each $\alpha_j$ and the position of every other source. This mass will be assumed to be positive, which constrains the value of the constant parameter $\kappa$.

In the above, we note that $r_{j}$ is defined by
\be \label{cosrj}
\cos r_j = \cos r_{Ij} \cos r + a_j\sin r_{Ij} \sin r,
\ee
for $j \neq I$, where the coordinate $r$ measures the radial distance from the $Ith$ black hole and $r_{Ij}$ is the distance from this black hole to the $jth$ black hole.  (Note that we can expand in powers of $r$, where $r=0$ is centred at the $Ith$ black hole, approximately close to $r=0$.) In addition, $a_k$ is defined by
\be \label{ak}
a_k \equiv \left[ \cos \theta_k \cos \theta - \cos\left(\phi_k-\phi\right)\sin \theta_k \sin \theta\right],
\ee
and $\left(r_k, \theta_k, \phi_k \right)$ indicates the location of the $kth$ black hole.

\newpage
\subsection{Eight regularly-spaced black hole masses}

We next study a specific configuration of masses. We consider eight identical mass  and equally spaced black hole (8BH) sources ($BH_1 - BH_8$) on a 3-sphere whose positions are displayed in Table \ref{table1}. In this table ($r$, $\theta$, $\phi$) are hyperspherical polar coordinates on the 3-sphere.
In particular, we note  that 
$\frac{\partial r_{1}}{\partial{r}} = - \frac{\partial r_{2}}{\partial{r}} =1$,
$\frac{\partial r_{1}}{\partial{\Phi}} =  \frac{\partial r_{2}}{\partial{\Phi}} = 
0$ and
$\frac{\partial r}{\partial{\chi}} \equiv -S[a_{\ell} \cos(r), \frac{\partial a_
{\ell}}{\partial{\Phi}} \sin(r)]$, where $\chi = \{r, \Phi\}$, $\Phi = \{\theta, 
\phi\}$, $\ell
= 3-8$, and $S(r) \equiv (1-a_{\ell}^2 \sin^2 (r))^{-1/2}$ so that
$\frac{\partial r_{\bar{\ell}+1}}{\partial{\chi}}=-
\frac{\partial r_{\bar{\ell}}}{\partial{\chi}}$ for $\bar{\ell} = 3,5,7$.

\begin{table}[h!]
\begin{center}
\begin{tabular}{|c|c|c|c|c|}
\hline

\bf{BH}   & \bf{($r_k$, $\theta_k$, $\phi_k$)}  &  $r_{1j}$ & $r_i$  & $a_{\ell}$  \\ 
\hline
$1$ & $\left(0, \frac{\pi}{2}, \frac{\pi}{2}\right)$  &  -- &  $r$   & -- \\
$2$   &  $\left(\pi, \frac{\pi}{2}, \frac{\pi}{2}\right)$  &  $\pi$  &   $\pi - r$  & -- \\
$3$  &  $\left( \frac{\pi}{2}, 0, \frac{\pi}{2}\right)$   &  $\pi/2$ &  $\cos(r_{\ell}) = a_{\ell} \sin(r)$   & $\cos\theta$ \\
$4$  &  $\left( \frac{\pi}{2}, \pi, \frac{\pi}{2}\right)$   & $\pi/2$ &  $"$   & $-a_3$ \\
$5$   &  $\left( \frac{\pi}{2}, \frac{\pi}{2}, 0\right)$   & $\pi/2$  &  $"$   &  $-\sin\theta \cos\phi$ \\
$6$  &  $\left( \frac{\pi}{2}, \frac{\pi}{2}, \pi\right)$   & $\pi/2$ &   $"$  &  $-a_5$ \\
$7$  &  $\left( \frac{\pi}{2}, \frac{\pi}{2}, \frac{\pi}{2}\right)$   &  $\pi/2$ &  $"$   & $-\sin\theta \sin\phi$ \\
$8$  &  $\left( \frac{\pi}{2}, \frac{\pi}{2},\frac{3 \pi}{2} \right)$   & $\pi/2$  &  $"$   &   $-a_7$\\
\hline
\end{tabular}
\end{center}

\caption{The positions of eight regularly arranged points on a 3-sphere in the 4D Euclidean embedding space with hyperspherical polar coordinates, where $I=1$, $i, j, k, \ell$ range from 1, 2, 3 to 8, respectively.
All definitions and
additional details are  described in the text.}

\label{table1}
\end{table}

\newpage

CCC qualitatively studied
the interesting question of whether the horizons of neighbouring black holes can touch \cite{bounce}. This would signal whether the black holes 
can retain their individual identity as the Universe bounces or whether they merge. 
In  \cite{bounce} the apparent horizon of each source
was located by determining the area of a sphere of constant  coordinate $r$ centred on the location of the black hole  \cite{DurkClif} (also see the Appendix). 
When the value of $r$ at the apparent horizon is greater than one half of the distance between the individual black hole sources, then we regard these black holes as having merged. 
The location of the apparent horizon for different values of $\kappa$ was displayed in Fig. 3 in CCC \cite{bounce}. In particular,
the horizons of two adjacent black holes touch at the  intersection of the two lines. This
formally occurs for $\kappa \simeq 5.1$; i.e.,  if 
$\kappa$ is greater than this value, then the black holes  are expected to merge prior to the expansion minimum.

CCC determined whether the scale factor of the cosmological region has a positive second time derivative corresponding to a bouncing cosmology.
Assuming that the cosmological region is approximately but not exactly (geometrically) spherical, it was found that it is plausible to infer that the cosmology bounces occurs when $\kappa  \gtrsim 1.2$ \cite{bounce}. In addition,
when $0 < \kappa  \lesssim  2.2$ the proper mass of each individual black hole is positive. However, the value of $m$ decreases as $\kappa$ increases, and at $\kappa \simeq 2.2$  it vanishes and  continues to decrease as $\kappa$  subsequently increases. Therefore, in the case of positive mass black holes, we necessarily have $\kappa \lesssim  2.2$.

By considering the positions of the horizons, the expansion of the cosmological scale factor and the property that the black hole masses are all positive, we thus have the following physical  bounds for the model parameter $\kappa$:
\be
\label{cond8}
1.2 \lesssim  \kappa \lesssim  2.2 \, .
\ee
If $2.2 \lesssim \kappa$, then the black holes have negative mass. If $\kappa \lesssim 1.2$, the magnitude of the scalar field  is then not sufficiently large to ensure a bounce in which the universe emerges into a subsequent expanding phase. Writing this in terms  of the radius of the apparent horizon, we have that
\be
\label{8bounds}
0.51 \lesssim  \Delta r_{\rm h} \lesssim  0.82 \, ,
\ee
where $\Delta r_{\rm h}$ represents the fractional distance on the conformal hypersphere that the horizon extends towards the midpoint between adjacent black holes. Therefore,
for all physical values of $\kappa$ we have that $\Delta r_{\rm h} < 1$.

\subsubsection{Comments}

Formally, the analysis presented in $CCC$ and summarized above is for the case $\kappa < 1$. But as noted above, for $\kappa < 1$ there is no bounce
(i.e., not physical), and when  $\kappa > 1$, for $\kappa \gtrsim 2.2$ the models have negative masses
(i.e., again not physical), and 
hence, strictly speaking, there is no black hole merger for all solutions in the physical regime. We need to consider the case $\kappa > 1$ explicitly. 
We recall that
bounces may occur for $\kappa$ greater than $\simeq 2.2$, but they will correspond to cosmological models with negative mass black holes. For the most part,  here we will assume that negative mass solutions are not physical.

We note that a  lot of intuition is broken in the BHL models studied here, which
can lead to difficulties in physical interpretation.
For example, the vacuum solution on a conformal hypersphere with one point mass doesn't exist; hence the model can not have a single BH limit (where all $\alpha_j\rightarrow 0$ except $\alpha_1=\alpha$). Moreover,
the masses can, in principle, be negative.  Most importantly, perhaps, the role of the Schwarzschild characteristic scale $R_S$ is not necessarily relevant. In particular, $m\sim 1$ may not be a good choice in the analysis regardless. From above, $m\equiv \alpha^2 B(\kappa)$, where $B$ depends on $\kappa$ (and where we assume a value for $\kappa$ such that $m>0$). Choosing $m$ effectively fixes $\alpha$. Alternatively, we could fix $\alpha$ so that $m$ is derived. However, as noted earlier, in our qualitative analysis here we shall keep $m=1$ in order to affect a more straightforward
comparison with the results of $CCC$ \cite{bounce}.

\subsubsection{$\kappa > 1$}

In particular, we need to consider the case $\kappa > 1$ separately and carefully.
If we assume that $K \equiv \sqrt{\kappa -1}$, then the physical region defined by (\ref{8bounds}) can be rewritten as:
\be
\label{cond8K}
0.45 \lesssim  K \lesssim  1.09 \, .
\ee
Again, the lower bound is to ensure a bounce  and the upper bound is to assure that
each black hole has a positive proper mass. Two problems are immediate. First, the assumption of spherical symmetry is even more  problematic in the
case  $\kappa > 1$. Second, we note that the region  $2.2 <  \kappa$ is unphysical. In particular, the critical merger value
$\kappa \simeq 5.1$ obtained in $CCC$ is not of physical interest.
However, we are interested in the
formal question of what happens as  $K$ increases (i.e., is there a critical value 
$K_c \sim \sqrt{5.1 -1}$ above which the individual black holes merge)?

\newpage

\subsection{The conformal factor for the case $ \kappa >1$}

Let us consider the case $ \kappa >1$.
We first define:  
$$K \equiv \sqrt{\kappa -1}.$$ We note that the earlier expressions are all  regular at $\kappa = 1$, where the trigonometric functions therein merely turn into hyperbolic ones
for $\kappa > 1$.

We obtain the conformal factor for a single point-like source at $r=0$
(which diverges at $r=0$, which is taken to be the location of the source).
To obtain a smooth geometry  at all other points, we require that the first derivative of $\Omega$ is single-valued at $r=\pi$. Hence we obtain the exact solution
\be
\label{cont333}
\Omega_i(r) = \alpha_i \frac{\sinh (K (\pi-r))}{\sinh (K \pi) \sin r} \, .
\ee

In the case of $N$ sources located at arbitrary positions on the hypersphere,  the corresponding solution can be  written as
\be
\label{cont}
\Omega (r, \theta, \phi) = \sum_{i=1}^N \alpha_i \frac{\sinh (K (\pi-r_i))}{\sinh (K \pi) \sin r_i} \, ,
\ee
where $r_i$  (identified by Eqs. \eqref{cosrj} and \eqref{ak}) is the radial coordinate, after rotating the hypersphere so that the $i$th source is positioned at $r_i=0$.

\subsubsection{Proper mass of sources}

Using the same procedure as described earlier (or as in \cite{bounce}), we obtain
the proper mass of the $i$th source as:
\be
m_i =  - \frac{2 \alpha_i^2 K}{ \tanh (K \pi)} + 2 \sum_{j\neq i} \alpha_i \alpha_j \frac{\sinh (K (\pi-r_{ij}))}{\sinh (K \pi) \sin r_{ij}} \, .
\ee
We remark that the mass of each individual source depends on every $\alpha_j$ and the position of every other black hole. As noted above, this mass can be positive or negative, depending on the value of the constants $\kappa$, $\alpha_i$ and $r_{ij}$.

\subsubsection{8BH configuration: case $ \kappa >1$}

We consider the  8BH case  with eight regularly-spaced masses and assume that
$\alpha_i = \alpha$, $m_i = m$. We then obtain:

\be
m= \frac{4 {\alpha}^2 \sinh(\frac{K\pi}{2})}{\sinh(K\pi)}
\left[3 - K \sinh(\frac{K\pi}{2}) \right] \, .
\ee
We have that $m=0$ when 
\be
3 - K \sinh \left[\frac{K\pi}{2} \right] = 0 \, ,
\ee
which is zero when $K \simeq 1.09$ (corresponding to $\kappa \simeq 2.2$). 
Notice that $-m {\alpha}^{-2}$ is positive and increases as $K$ increases.

\newpage

\section{Evolution of instantaneously-static models}

We next present perturbative solutions of the constraint Eq. (\ref{constraint}) that can then be utilized to construct exact time evolving solutions of the EFEs. The
resulting 4D spacetimes model physical realisations in which multiple distinct black holes persist through a cosmological bounce. 

In order to study the complete dynamics of the models it is necessary to solve the full EFEs and not simply the conservation Eqs. and the constraint Eqs. 
We choose coordinates in order that the full 4D spacetime is represented by the metric line-element:
\be
\label{dynmetric}
ds^2 = -A^2(t,x^{\gamma}) dt^2 +  h_{\alpha \beta} (t,x^{\gamma}) dx^{\alpha} dx^{\beta}.
\ee
An evolving solution would correspond to such a metric which satisfies both the constraint and evolution Eqs. For a dynamical model that bounces at the time-symmetic surface at $t=t_0$, the 3-metric is:
\be
\label{hypmetric}
h_{\alpha \beta} (t_0,x^{\gamma}) dx^{\alpha} dx^{\beta} = \Omega^4(x^{\gamma})d\tilde{s}^2
\equiv \Omega^4 \tilde{h}_{\alpha \beta} dx^{\alpha} dx^{\beta}
 \, .
\ee
An exact {\em{spacetime}} can then be obtained by solving the evolution Eqs.
\bea
\mathcal{L}_t h_{\alpha \beta} &=& -2 A K_{\alpha \beta}
\eea
and
\bea
\mathcal{L}_t K_{\alpha \beta} &=& - D_{\alpha} D_{\beta} A + A \left( \mathcal{R}_{\alpha \beta} - 2 K_{\alpha \gamma} K^{\gamma}_{\phantom{\gamma} \beta} + K \, K_{\alpha \beta} \right) \nonumber \\ &&+ \frac{A}{2} \left( p^{\varphi} - \rho^{\varphi} - 2 \Pi^{\varphi}_{\alpha \beta} - \frac{2}{3}\rho^{\gamma} \right) h_{\alpha \beta} \, ,
\eea
where $t^{a} \equiv A u^{a}$, {and} $\Pi^{\varphi}_{a b} = (h_a^{\phantom{a} c} h_b^{\phantom{b} d} - \frac{1}{3} h_{ab} h^{cd}) T^{\varphi}_{cd}$ is the effective anisotropic stress of the scalar field. Because the initial value problem is well-posed, 
the specification of suitable initial values for both the metric and the extrinsic curvature guarantees a unique time evolving solution. 
We will next obtain approximate solutions.

\subsection{Perturbative time-evolving solutions}
Assume that the metric functions ($A$, $h_{\alpha\beta}$ in (\ref{dynmetric})) are smooth at the (symmetric) bounce at $t=0$ and can be expanded in powers of $t$. We also expand $T^\varphi_{ab}$, given by Eq. (2), in powers of $t$, and define the time-like vector field $\frac{1}{A}\frac{\partial}{\partial t}$ so that a time derivative in the 1+3 split is defined by $\dot{\chi}=\frac{1}{A}\frac{\partial\chi}{\partial t}$.

At $t=0$, we have that
\begin{equation}
K_{ij}=0, \hspace{2mm} \tilde{D_a}\varphi=0, \hspace{2mm} \dot{\Pi}=0, \hspace{2mm} V^{\prime}=0.
\end{equation}
In addition, we recall that
\be
\tilde{D}^2\Omega=\kappa\Omega
\ee
and (at $t$=0)
\be
\Pi^2=\dot{\varphi}^2=\frac{\lambda^2}{\Omega^4} \hspace{2mm} ; \hspace{2mm} \lambda^2 \equiv \frac{8}{\mu}\left(\frac{3}{4}-\kappa\right)
\ee
so that $\kappa>\frac{3}{4}$ for $\mu=-1$. Also
\be
h^0_{\alpha\beta}\equiv h_{\alpha\beta}\left(t=0,x^\gamma\right)=\Omega^4\tilde{h}^0_{\alpha\beta},
\ee
where
\be
\tilde{h}^0_{\alpha\beta}=diag\left[1, \sin^2r, \sin^2r \sin^2\theta\right].
\ee
From  $K_{\alpha\beta}=0$ and the EFE, we find that by a translation of $r$ and $t$ we can always set $A=1$, and all linear terms in $h_{\alpha\beta}$ vanish. Hence we have that:
\be
h_{\alpha\beta}=h^0_{\alpha\beta}+h^2_{\alpha\beta}t^2+o(t^3),
\ee
where $h^0_{\alpha\beta}$ is defined by (49) and (50). We only need $h_{\alpha\beta}$ to second order in order to calculate the Riemann tensor at $t=0$, but we need the metric to order third order in order to compute the covariant derivative of the Riemann tensor at $t=0$. In the absence of radiation (i.e., a scalar field source only), the bounce can be assumed to be symmetric and hence the next contributions to the metric will be $o(t^4)$. But radiation can be included at $o(t^3)$ self-consistently to obtain an evolving non-symmetric (about the bounce) solution.

We can compute the inverse metric, where $g^{00}=-1$, and
\be
h^{\alpha\beta}=h^{\alpha\beta}_0+o(t^2),
\ee
where
\be
h^{\alpha\beta}_0=\Omega^{-4}diag\left[1,\frac{1}{\sin^2r},\frac{1}{\sin^2r\sin^2\theta}\right]\equiv\Omega^{-4}\tilde{h}^{\alpha\beta}_0,
\ee
and we only need $h^{\alpha\beta}$ to zeroth-order, $h^{\alpha\beta}_0,$ for the computations here.
The field Eqs. (44) and (45) (for $K_{\alpha\beta}=-h_\alpha^\gamma h_\beta^\delta\nabla_\gamma u_\delta,$ $\mathcal{L}_th_{\alpha\beta}=-2K_{\alpha\beta}$) at $t=0$ then yield
\be
K_{\alpha\beta,t}=-h^2_{\alpha\beta}.
\ee

Now, we can explicitly compute the non-trivial connection coefficients to lowest order:
\begin{equation}
\Gamma^0_{\alpha\beta}=h^2_{\alpha\beta}t, \hspace{2mm} \Gamma^\alpha_{0\beta}=\Omega^{-4}h^2_{\alpha\beta}t, \hspace{2mm} \Gamma^\alpha_{\beta\gamma}=\tilde{\Gamma}^\alpha_{\beta\gamma} + o(t^2),
\end{equation}
and similarly for the partial derivatives of the connection coefficients. We can then compute the components of the Riemann tensor. In particular, 
from the Eqs. (45) we find that
\be
h^2_{\alpha\beta}=-R^0_{\alpha\beta}
\ee
(where the 3D Ricci tensor $R^0_{\alpha\beta}$ is calculated explicitly from $h^0_{\alpha\beta}$ and contains derivatives of $\Omega$, which can be simplified using (47)). 

Expansions for $R_{\alpha\beta}, R_{\alpha\beta\gamma\delta}, W_{\alpha\beta\gamma\delta}$, and their first covariant derivatives in terms of $R^0_{\alpha\beta}, R^0_{\alpha\beta\gamma\delta}, W^0_{\alpha\beta\gamma\delta}$, and higher order terms (higher powers of $t$) in terms of $\Omega$ and its derivatives, can be explicitly computed. In particular,
\begin{align}
\begin{split}
&R_{\alpha\beta\mu\gamma} =\tilde{R}_{\alpha\beta\mu\gamma}+o(t^2), \\
&R^0_{\beta\mu\gamma} \sim o(t), \\
&R^0_{\beta0\gamma} =-\frac{1}{2}h^2_{\beta\gamma}=\frac{1}{2}R^0_{\beta\gamma}+o(t^2),
\end{split}
\end{align}
using Eq. (56). We write:
$$\varphi=\phi_o+\Phi_1(r)t+o(t^3),$$
where $\phi_o=const$, there is no $o(t^2)$ term since the acceleration is zero, and from (46):
\begin{align}
\begin{split}
\Phi_1(r)&=\frac{\lambda}{\Omega^2}\equiv\Phi(r), \\
V(\varphi)&=0+V_1(r)t+V_2(r)t^2+o(t^3),
\end{split}
\end{align}
where $V_0=0$ since $V^{\prime}=0$ at $t=0$. We note that $\rho^{\varphi}=0$ at $t=0$, consistent with Eqs. (2) and (3) above. The potential $V(\varphi)$ can be determined from the conservation Eq. (2) for the scalar field (if $\phi_0\neq0$, $V_2=0$, else if $\phi_0=0,$ $V_2\neq0$).

We can compute $T_{ab}$ explicitly, where $T_{ab}=T_{ab}^\varphi$ is defined by (2). Note that (dropping the index 1 on $\Phi_1(r)$)
\begin{equation}
	\begin{split}
	\dot{\varphi} &=\Phi+o(t^2), \\
	D_a\varphi \equiv \psi_a &=t\Phi_{,\alpha}+o(t^3), \\
	\dot{\Pi} &=o(t);
	\end{split}
\end{equation}
we then have that
\begin{equation}
	\begin{split}
	\rho^\varphi = p^\varphi &=\frac{1}{2}\Phi^2, \\
	D^a \varphi D_a \varphi &\sim o(t^2), \\
	\nabla_\alpha \varphi \nabla_\beta \varphi &= \Phi_{,\alpha} \Phi_{,\beta} t^2 + o(t^2), \\
	\nabla^a \nabla_a \varphi &= -\Phi^2 + o(t^2).
	\end{split}
\end{equation}
Therefore,
\begin{equation}
	\begin{split}
	T_{00} &= \frac{1}{2} \Phi^2 + o(t), \\
	T_{0\alpha} &= \Phi\Phi_{,\alpha} t + o(t^2), \\
	T_{\alpha\beta} &= \frac{1}{2} h^0_{\alpha\beta} \Phi^2 + o(t), \\
	T = g^{ab} T_{ab} &= \Phi^2 + o(t).
	\end{split}
\end{equation}
We can obtain the higher order terms for $V$ by perturbatively solving the conservation Eq. (3) (to lowest order $\dot{\varphi} = \Phi(r) + \chi(r)t^2$; $V(\varphi)=\overline{V}_2 \varphi^2$; yielding Eqs. for $\chi$ and $\overline{V}_2$). In particular, Eq. (3) implies that $\phi_0 V_2 = 0$.

Explicitly, the 4D Ricci tensor at $t=0$ can be written as:
\begin{equation}
	^4R_{00} = \prescript{3}{}{R}, \hspace{2mm} ^4R_{0\alpha} = 0, \hspace{2mm} ^4R_{\alpha\beta} = 0, \hspace{2mm} ^4R = - \prescript{3}{}{R}.
\end{equation}
The field Eqs.
\begin{equation}
	R_{ab} - \frac{1}{2} g_{ab} = -T_{ab}
\end{equation}
yield $^4R = \prescript{4}{}{T}$ $\left(\prescript{4}{}{T} = - \prescript{3}{}{R}\right)$. Clearly, using Eqs. (59) -- (61), the EFEs are satisfied. Note that $T_{ab;c} g^{bc} = 0$ to lowest order.

\subsubsection{Curvature invariants}

We can use the tensor $T_{ab}$ to compute curvature invariants. Note that, to lowest order, $T = \Phi^2,$ $T_{ab}T^{ab} = \Phi^4$. Defining the trace-free tensor
\be
	S_{ab} = T_{ab} - \frac{1}{4}g_{ab}T,
\ee
which is related to the trace-free Ricci tensor, we have that
\be
	S_{00} = \frac{3}{4}\Phi^2, \hspace{2mm} S_{\alpha\beta}=\frac{1}{4}h_{\alpha\beta}\Phi^2; \hspace{2mm} S=0
\ee
and
\be
	\begin{split}
\tensor{S}{_a^b} \tensor{S}{_b^a} & = \frac{3}{4}\Phi^4, \\
\tensor{S}{_a^b} \tensor{S}{_b^c} \tensor{S}{_c^a} &= -\frac{3}{8} \Phi^6, \\
\tensor{S}{_a^b} \tensor{S}{_b^c} \tensor{S}{_c^d} \tensor{S}{_d^a} &= \frac{21}{64} \Phi^8,
	\end{split}
\ee
and the type {\bf II/D} condition \cite{PLB} becomes
\be
\label{DT}
	^4D_T = {a_0}^2\Phi^{24} = 0,
\ee
where ${a_0}^2$ is a positive non-zero constant.

In addition, computing $T_{ab;c}$ to $o(t^0)$:
\begin{align}
\begin{split}
T_{00;c} &= \Phi\Phi_{,\gamma}\delta^\gamma_c, \\
T_{0\alpha;c} &= \Phi\Phi_{,\alpha}\delta^0_c, \\
T_{\alpha\beta;0} &= 0, \\
T_{\alpha\beta;\delta} &= h^0_{\alpha\beta} \Phi\Phi_{,\delta}
\end{split}
\end{align}
(noting that the higher order terms in $T_{\alpha\beta;0}$ contain appropriate terms  involving the potential), so that:
\be
T_{ab;c}T^{ab;c} = 6 \Phi^2\Phi_{,\alpha} \Phi_{,\beta} h_0^{\alpha\beta}
\ee
and the type {\bf II/D} condition for the covariant derivative of $T_{ab}$ (or the trace-free $S_{ab}$),  $\prescript{4}{}{D}_{\nabla T} =0$, becomes 
\be
\label{DDT}
\Phi^2\Phi_{,\alpha}\Phi_{,\beta}h^{\alpha\beta}_0 = 0.
\ee
Note that all contractions of the Weyl tensor are proportional to $\left(\Phi^2\right)^{n_1}$, and all contractions of the covariant derivative of the Weyl tensor are proportional to $\left(\Phi^2\Phi_{,\alpha}\Phi^{,\alpha}\right)^{n_2}$, for positive integers $n_1$ and $n_2$. The GH is identified through Eqs. (\ref{DT}) and (\ref{DDT}).

Therefore, effectively the type {\bf II/D} condition for $T_{ab}$ yields
\be
\Phi=0,
\ee
and for $\nabla T_{ab}$ (in the spherical limit):
\be
\frac{d\Phi}{dr}=0.
\ee
The first condition (67) or (71) enables the proper masses to be identified. The second type {\bf II/D} condition (70), $\Omega_{,\alpha} \Omega^{,\alpha} = 0$, allows us to identify the GH. This condition is analogous to the condition $\partial_r\left(E^{11}\right)=0$ in \cite{DurkClif} using the so-called Weyl-tensor method (see the Appendix). Note that at $t=0$, $E_{\mu\nu}=\prescript{3}{}{R}_{\mu\nu}$ and $H_{\mu\nu}=0$ (the electric and magnetic parts of the Weyl tensor, respectively). The type {\bf II/D} condition for the Weyl tensor is $I^3=\lambda J^2$, where $I\equiv\frac{1}{2}\left(E_{ab}E^{ab}-H_{ab}H^{ab} + iE_{ab}B^{ab}\right)$ (and similarly for $J$) \cite{kramer}.

The analysis can be repeated in flat space and higher dimensions (see Appendix); however, we can only use the GH as above to identify the black hole horizons in higher dimensions.

\newpage

\subsection{Example: hyperspherical black holes} As an example, for the hyperspherical black holes cosmology studied earlier, $\Omega(r,\theta,\phi)$ is given by Eq. (26). Note that
for small $r$ $(j \neq I)$:
\be
r_j = r_{Ij} - a_j r + o(r^2).
\ee
From earlier $\prescript{4}{}{D}_T \propto \Phi^{24}$ and $\Phi=\frac{\lambda}{\Omega^2}$. As in \cite{DurkClif}, the expression $\Omega$ is used to identify the proper masses of the black holes. From $\prescript{4}{}{D}_{\nabla T} = 0$, we obtain $\Omega_{,\alpha}\Omega_{,\beta} h^{\alpha\beta}_0 = 0$, from which we determine the GH. Thus, we find that
\be
\Omega(r,\theta,\phi) = \frac{\alpha_I}{r} + \frac{m_I}{2\alpha_{I}} + r\left\{ \left(\frac{\kappa}{2} - \frac{1}{3}\right) \alpha_I + C_I \right\} +o(r^2),
\ee
where the proper masses $m_I$ are defined by
\be
\frac{m_I}{2\alpha_I} = \frac{-\alpha_I \sqrt{1-\kappa}}{{\tan(\sqrt{1-\kappa}~\pi)}} + B_I,
\ee
and where
\begin{align}
B_I &\equiv \sum_{j \neq I}B_{I_j} = \sum_{j \neq I} \frac{\alpha_j}{\sin r_{Ij}}\left[\cos \left(\sqrt{1-\kappa}~r_{Ij}\right) - \cot \left(\sqrt{1-\kappa}~\pi\right) \sin \left(\sqrt{1-\kappa}~r_{Ij}\right)\right], \\
C_I &= \sum_{j \neq I} a_j \left[B_{I_j} \cot r_{Ij} - B_{I_j} \sqrt{1-\kappa}~\cot \left(\sqrt{1-\kappa}~r_{Ij}\right) + \frac{\alpha_j \sqrt{1-\kappa}}{\sin r_{Ij} \sin\left(\sqrt{1-\kappa}~r_{Ij}\right)}\right].
\end{align}
Note that (for small $r$):
\be
\Omega_{,r} = \frac{-\alpha_I}{r^2} + \left\{ (3\kappa-2)\frac{\alpha_I}{6} + C_I \right\},
\ee
and the GH, $r_{gh}$, is defined when this vanishes:
\be \label{78}
r_{gh} = \left[ \frac{(3\kappa-2)}{6} + \frac{C_I}{\alpha_I} \right]^{-\frac{1}{2}}.
\ee

\subsubsection{8BH case}

First we consider the case  $\kappa < 1$ under the spherical assumption, with  equal masses $m_i = m$ ($\alpha_i = \alpha$), where we shall take $m=1$ as in CCC. We also assume a single cell (where the antipodal points $BH_1$ and $BH_2$ are not identified, and there are restrictions on $\theta, \phi$ within this single cell). Note that we need to consider the case $\kappa > 1$ separately and carefully (see later); in particular, in this latter case we need to relax the assumption of spherical symmetry and we need to utilize the GH.

In this specific example we consider a regular 8BH lattice and explicitly determine the proper masses and location of the GH and compare them with results using the Weyl tensor method (see the Appendix). The qualitative agreement further motivates the use of the GH to identify the horizon of the black holes.
We note that
Eq. \eqref{78} is only useful heuristically. We have assumed that $\frac{3}{4} < \kappa < 1$ here and looked at the spherically symmetric approximation for small $r$. Eq. \eqref{78} is not valid for larger $r$, and so cannot be used to estimate $r_{gh}$ (but we can adjust the relative size by decreasing $m$ (or $\alpha$)).
To try and obtain analytical results to confirm the qualitative picture we
can choose $\kappa = \frac{15}{16}$ $\left(\sqrt{1-\kappa} = \frac{1}{4}\right)$.

Within the approximation here,
$$\frac{C_i}{\alpha} = c^2\sin\theta\sin\phi,$$
where $c^2 \equiv (1-\kappa)\cos(\sqrt{1-\kappa}~\pi)\sin^{-2}(\sqrt{1-\kappa}~\pi) > 0$ for $\frac{3}{4}<\kappa<1$. We can consider the range of values $0<\theta<\frac{\pi}{2}$, $0<\phi<\frac{\pi}{2}$ so that $C_i$ is positive (other ranges can be obtained by symmetries for the symmetric 8BH configuration).
For $\frac{C_1}{\alpha}\approx 0$, $r_{gh}^{-2} \cong \frac{3\kappa -2}{6}$, so that qualitatively as $\kappa$ increases $r_{gh}$ is decreasing, which looks promising. Note that $C_1>0$, so that $r_{gh}$ always decreases relative to the``pure" spherically symmetric case. Also note that as $\alpha$ decreases, $r_{gh}$ decreases.

If the black holes merge, they do so first along the lines joining the centres of the black holes: Let $^jC_1$ be the value of $C_1$ along the lines joining $BH_1$ and $BH_j$, where
$$^j C_1 = \alpha c^2(1,0,0,0,0,1,-1).$$
Therefore, we can estimate the effect of $C_1$ in these critical directions.

We also  need to redo the computations for $r$ not small (e.g., $r\simeq \frac{1}{2}r_{ij} \sim \frac{\pi}{4}$), and especially without any approximation for $r$ (to find all roots of $\Omega_{,\alpha}\Omega^{,\alpha} = 0$).  We note that
\begin{equation}
\sinh\left(nK\pi\right) = \frac{1}{2}e^{nK\pi}\left(1-e^{-2nK\pi}\right), 
\end{equation}
and we can neglect the second term (i.e., 
$\sinh\left(nK\pi\right) \simeq \frac{1}{2}e^{nK\pi} \simeq \cosh\left(nK\pi\right)$)
since  in the large K approximation  $ e^{-2nK\pi} < 10^{-2}$ for $nK > \frac{1}{3}$.

\newpage

\section{8BH: The Case $\kappa>1$}

We assume a single cell 8BH configuration.  As noted above, for our qualitative analysis here we shall keep $m=1$ in order to be able to compare with the results of $CCC$ \cite{bounce}. We take $r=0$ to be centred on BH$_1$ ($I=1$).  We consider equal masses ($m_i=m=1$) and equal $\alpha_i=\alpha$, where all $r_{Ij}=\frac{\pi}{2}$, except $r_{I2}=\pi$.

The physically viable range for $K$ is $0.45 \lesssim  K \lesssim  1.09$. 
We shall assume that $K\lesssim  1.09$ ($\kappa \lesssim  2.2$) in order for the masses to be positive. Strictly speaking, the analysis in $CCC$ is not valid for $\kappa \gtrsim 2.2$. However, we can formally investigate what happens for large $\kappa$ here. Also, formally there is a bounce for $K\gtrsim 0.45$ ($\kappa \gtrsim 1.2$).
Note that it was found in the analysis of $CCC$ that the black holes do not merge in this physically acceptable range.

This implies that the ranges for $\theta$, $\phi$ may be restricted in order for the region (and the GH in particular) to lie within the interior of the cell.
We shall choose BH locations and ranges for the angular values below appropriate to the applications (and appeal to the symmetries of the configuration otherwise). We are most interested in the directions (angular values) joining two black holes.

The  analysis in the case $\kappa<1$ discussed above is really only applicable for small $r$. The spherically symmetric limit and use of an apparent horizon is even less appropriate in the large K limit (LKA).  Indeed, the GH will be angular dependent. Therefore, we 
need to redo the analysis in the case $\kappa>1$.

We also have that
\begin{align}
\Omega = &\sum^8_{i=1}\alpha_i \frac{\sinh \left(K(\pi-r_i)\right)}{\sinh (K\pi) \sin r_i}
\end{align}
and, using  $\alpha_i \equiv \alpha$, hence
	\begin{align}\label{star}
		\alpha^{-1} \sinh (K\pi) \hspace{1mm} \Omega &= \frac{\sinh\left[K(\pi-r)\right]}{\sin r} + \frac{\sinh(Kr)}{\sin r} + \sum^8_{\ell=3} \frac{\sinh (K(\pi-r_\ell))}{\sin r_\ell}\\ \nonumber
		&\equiv F(r) + G(r,\theta,\phi).
	\end{align}
	We have that
	\begin{align}\label{daggar1}
		\Theta &\equiv \alpha^{-2}\sinh^2(K\pi) \left[\Omega_{,i}\Omega_{,j}\right] h^{ij} = \left[ F_{,r} + G_{,r} \right]^2 + \frac{1}{\sin^2 r}\left[ G_{,\theta}\right]^2 + \frac{1}{\sin^2r \cos^2\theta} \left[G_{,\phi}^2\right],
	\end{align}
where
	\begin{align}
		F(r) &\equiv \frac{1}{\sin r}\left[\sinh (K\pi) \cosh (Kr) - \cosh (K\pi) \sinh(Kr) + \sinh (Kr)\right],\\ \nonumber
		A(r) &\equiv \left[F(r)\right]_{,r} = \frac{-\cos r}{\sin^2r}\left[\sinh(K\pi)\cosh(Kr)-\cosh(K\pi)\sinh(Kr) + \sinh(Kr)\right]\\ 
		& \hspace{17mm} + \frac{K}{\sin r} \left[\sinh (K\pi) \sinh(Kr) - \cosh(K\pi)\cosh(Kr) + \cosh(Kr)\right].
	\end{align}
We rewrite (82) and \eqref{daggar1} using $\ell=3 - 8$ as:
	\begin{align}
		&\alpha^{-1}\sinh(K\pi)\Omega = F(r) + G(r,\theta,\phi),\\
		&G(r,\theta,\phi)=\sum_{\ell=3}^{8}\frac{\sinh(K(\pi-r_{\ell}))}{\sin r_{\ell}} \equiv \sum^8_{\ell=3}G_{\ell}.
	\end{align}
We note that
	\begin{equation}
		F(r)=\frac{1}{\sin r}\left[ \sinh(K\pi)\cosh(Kr) + \sinh(Kr)(1-\cosh(K\pi))\right],
	\end{equation}
where $F_{,r}\equiv A(r)$ has no angular ($\theta, \phi$) dependence.\\

For each $\ell$ and $\chi \equiv [r,\theta,\phi]$, we have that
	\begin{align}
		\tensor{G}{^\ell_{,\chi}} = \frac{\partial r_{\ell}}{\partial{\chi}}&\left\{ \frac{-\cos(r_\ell)}{\sin^2(r_\ell)} \left[\sinh(K\pi)\cosh(Kr_\ell) - \cosh(K\pi)\sinh(Kr_\ell)\right]\right. \nonumber \\
		&+ \left. \frac{K}{\sin (r_\ell)} \left( \sinh(K\pi)\sinh(Kr_\ell)-\cosh(K\pi)\cosh(Kr_\ell)\right)\right\}.
	\end{align}
We define
	\begin{equation}
G = \sum^8_{\ell =3}G_{\ell}=\sum_{\bar\ell=3,5,7}\big(G_{\bar\ell}+G_{\bar\ell +1}\big) \equiv \sum^{\bar3}_{L=\bar1} G_L,
	\end{equation}
where each $G_L(r_L) \equiv G_{\bar\ell} + G_{\bar\ell +1}$;
$L=\bar1$ $(\bar\ell=3)$, $L=\bar2$ $(\bar\ell=5)$, $L=\bar3$ $(\bar\ell = 7)$.
Now, for each dipodal pair $\bar\ell$ and $\bar\ell+1$,  by a direct computation we have that $r_{\bar\ell+1}=\pi-r_{\ell}$. 
We note that explicitly
	\begin{equation}
		A(r) = -\frac{1}{\sin r} \bigg[ \cot r \big(\sinh K(\pi-r) + \sinh(Kr)\big) + K\big(\cosh K(\pi-r)\big) - \cosh (Kr))\bigg],
	\end{equation}
so that immediately we have that $A(\pi-r)=-A(r)$ (and hence that $A(\frac{\pi}{2})=0$).

Hence we obtain the exact result:
	\begin{equation}
		\big(G_L\big)_{,\chi} = \big(G_L(r_L)\big)_{,r_L}\frac{\partial r_L}{\partial\chi}; ~~G_{,\chi} = \sum^{\bar3}_{L=\bar1}S_L \tensor{b}{_L^\chi}A(r_L),
	\end{equation}
where $A(r)$ is defined above, $\cos r_L = a_L\sin r$, and
$$S_L \equiv -\frac{1}{\big(1-a_L^2\sin^2r\big)^\frac{1}{2}},$$
	\[ b^\chi_L \equiv \begin{cases}
			a_L\sin r & \chi=r\\
			\frac{\partial a_L}{\partial \chi} \sin r & \chi=\theta,\phi
			\end{cases}.
	\]
We recall that $\Theta$ is given by \eqref{daggar1}.

\subsubsection{Approximation method}

Let $r=r_0+x$ (for some $r_0$; e.g., $r_0=r_{gh}$), where $x$ is small (so we can expand about $x=0$). By definition,
	\begin{equation}
		\cos r_{\bar i} \equiv \bar A_{\bar i}\sin r = \bar A_{\bar i} \sin r_0 + \bar A_{\bar i} \cos (r_0) x \equiv \bar A + \bar B x, \nonumber
	\end{equation}
and so
\begin{equation} \label{doubledaggar}
r_{\bar i} = \arccos(\bar A + \bar B x) = \arccos(\bar A) - \frac{\bar B}{\sqrt{1-\bar A ^2}}x \equiv R_{\bar i} + P_{\bar i} x + o(x^2),
\end{equation}
where
	\begin{equation} \label{doublestar}
		R_{\bar i} \equiv \arccos \big(\bar A_{\bar i} \sin r_0 \big); ~~
		P_{\bar i} \equiv \frac{- \bar A_{\bar i} \cos r_{0}}{\sqrt{1-\bar A_{\bar i}^2 \sin^2r_0}}.
	\end{equation}
Note that for small $x$:
	\begin{equation}
	\begin{split}
&A\big(\alpha +\beta x\big) = A\big(\alpha\big) \\
		 &+ \beta x\left\{\frac{\cos\alpha}{\sin^2\alpha}\bigg[\big[\tan\alpha+2\cot\alpha\big] \big(\sinh K(\pi-\alpha)+\sinh K\alpha\big)-K\big(\cosh K\alpha-\cosh K(\pi-\alpha)\big)\bigg]\right. \\
		&+\left. \frac{K}{\sin r}\bigg[ -\cot \alpha\big(-\cosh K(\pi-\alpha)+\cosh K\alpha \big)+K\big(\sinh K(\pi-\alpha) + K\sinh K\alpha\big)\bigg]\right\}.
	\end{split} \label{dagger}
	\end{equation}

\noindent
Here we assume that all $a_L$ are positive and $0\leq\theta\leq\frac{\pi}{2}$, $0\leq\phi\leq\frac{\pi}{2}$, to ensure that $r_L < \frac{\pi}{2}$ (other regions are obtained by ``symmetry arguments").

There are 3 ``pairs'': $ L = \bar 1, \bar 2, \bar 3$, where
$\cos r_{\bar L} = a_{\bar L} \sin r.$
For the 3 dipodal pairs we take
	\begin{align}
		0&< r\leq \frac{\pi}{4} \\
		0&< a_{\ell}\sin r < \frac{1}{\sqrt 2} \\
		0&< \cos r_{\bar L} <\frac{1}{\sqrt 2}\\
		\frac{\pi}{4} &< r_{\bar L} < \frac{1}{\sqrt{2}}
	\end{align}
For example, in general we take
	\begin{align}
		&a_{\bar 1}=\cos \theta \hspace{3mm} (= a_3),\\
		&a_{\bar 2}=\sin\theta\cos\phi \hspace{3mm} (=-a_5),\\
		&a_{\bar 3}=\sin\theta\sin\phi \hspace{3mm} (=-a_7),
	\end{align}
	
\vspace{0.3cm}  	

\noindent so that
\begin{table}[h!]
  \begin{center}
    \label{tab:table1}
    \begin{tabular}{r|c|c|c}
       & $a_{L}$ & $\frac{\partial a_L}{\partial \theta}$ & $\frac{\partial a_L}{\partial \phi}$\\
      \hline
      $\bar 1$ & $\cos\theta$ & $-\sin\theta$ & $0$ \\
      \hspace{2mm} & \hspace{2mm} & \hspace{2mm} & \hspace{2mm}\\
      $\bar 2$ & $\sin\theta\cos\phi$ & $\cos\theta\cos\phi$ & $-\sin\theta\sin\phi$ \\
      \hspace{2mm} & \hspace{2mm} & \hspace{2mm} & \hspace{2mm}\\
      $\bar 3$ & $\sin\theta\sin\phi$ & $\cos\theta\sin\phi$ & $\sin\theta\cos\phi$ \\
    \end{tabular}
  \end{center}
\end{table}\\

\vspace{0.3cm}  

\noindent
In a specific example we may change coordinates (i.e., the  choice of $a_{\bar 1},$ $a_{\bar 2},$ $a_{\bar 3}$, depending on whether BH$_3$/BH$_4$, BH$_5$/BH$_6$, BH$_7$/BH$_8$ are chosen) to ensure that $0<\theta<\frac{\pi}{2}$ and $0<\phi<\frac{\pi}{2}$ and that all terms are thus well-defined in that specific example.

\subsubsection{Large K Approximation}

Note that in most calculations with $K \gtrsim 1$ $(0<a_{\ell}<1)$, $r<\frac{\pi}{4}$, $\frac{\pi}{4}<r_L<\frac{\pi}{2}$, and we can utilize a``large K" approximation (LKA):
	$$\sinh nK\pi  \simeq \cosh nK\pi  \simeq \frac{1}{2}e^{nK\pi},$$
where the amplitude of the neglected terms vary depending on $n$ (and $K$), but in most applications of interest with $K > 1$ the neglected terms are less than $1\%$, relative to the dominant terms. Note that special care must be taken in which terms to neglect in the case $r\sim\frac{\pi}{4}$, where various terms with exponential factors such as $e^{\frac{\pi}{2}}$ and $e^{\pi-2r}$ are comparable. In such cases analytic computations for precise values of $\theta$ and $\phi$ are preferable. For example, within the LKA approximation and for $\chi=r$ we obtain
	$$G_{,r}=\sum^{\bar 3}_{L=\bar 1} \frac{-a_L\cos r}{(1-a_L^2\sin r)^{\frac{1}{2}}}A\left(r_L\right),$$
where
	\begin{align} \nonumber
		A\left(r_L\right)&=\frac{-\cos r_L}{\sin^2r_L}\bigg[\sinh K(\pi-r_L) + \sinh Kr_L\bigg]\\ \nonumber
		& + \frac{K}{\sin r_L}\bigg[ -\cosh K(\pi-r_L) + \cosh Kr_L\bigg]\\
	&\simeq \frac{-\frac{1}{2}e^{K(\pi - r_L)}}{\sin r_L}\bigg[\big(K+\cot r_L\big) + o\left(e^{-K(\pi -2r_L)}\right)\bigg].
	\end{align}

Therefore, in the LKA the correction terms can be neglected. Indeed, for $r<\frac{\pi}{2}$ and $r_L<\frac{\pi}{4}$,
	$$\frac{A\left(r_L\right)}{A(r)} \sim e^{K\left(r-r_L\right)}.$$
For $r\simeq r_0 < \frac{\pi}{4}$, $r_L \simeq \arccos(\bar A_L \sin r_0)$ ($|\bar A_L|<1$), so that $r_L>\frac{\pi}{4}>r_0$ (i.e., $r-r_L<0$), and as $r_0$ decreases $r_L-r$ increases. Hence the terms $A(r_L)$ become increasingly negligible relative to $A(r)$.
As noted above, and as we shall see below, $r\sim \frac{\pi}{4}$ must be treated separately.

We also note that for small $r$, the $\frac{1}{\sin r}$ terms in \eqref{daggar1} and the $\sin r$ factors in $G_{,\theta}$ and $G_{,\phi}$ cancel and there are 
consequently no ``degeneracies" (i.e., the $r\to 0$ limit of Eq. \eqref{daggar1} is well-defined). However, the computations here are not valid for small $r$, and we utilize the qualitative analysis described earlier.
In fact, in almost all cases of interest we can show that the $G_{,\theta}$ and $G_{,\phi}$ terms in \eqref{daggar1} can be neglected.

\subsubsection{Comments}

We have obtained an analytical expression for $\Theta$. We now wish to examine the roots, $r=r_{gh}$, of $\Theta=0$. We can study this for a particular value for $K$. Since the physical range for $K$ is $0.45<K \lesssim  1.09$, interesting values for $K$ may include $K=\frac{2}{\pi}\sim 0.63$, $K=1$, and $K=\frac{4}{\pi}$. 
In the following applications we can consider the LKA approximation (where appropriate).
We might also consider particular ranges for $r$. For example, we considered the small $r$ limit earlier (where $r_{gh}$ decreases with increasing $K$). However, $r_{gh}$ is not expected to be very small. We can do detailed approximate calculations for $r \simeq \frac{\pi}{4}$.
Numerical plots are often useful.
It is also of interest to study $r_{gh}$ for particular and important angular values (e.g., $\theta=0$ and $\theta=\frac{\pi}{2}$), or for a range of small angular values (of $\theta$, for example, about $\theta=0$ for $\phi=\frac{\pi}{2}$).

\newpage

\subsection{The case $\theta=0$} 

Here we have that $a_{\bar 1}=1$, $a_{\bar 2}=a_{\bar 3}=0$,
	\begin{align}
		b^r_{\bar 1} &= \cos r,\\
		b^r_{\bar 2} &= b^r_{\bar 3}=0,\\
		b^{\phi}_{L}&=0 \hspace{3mm}\text{ (all L),}\\
		b^{\theta}_{\bar 1}&=0, \hspace{3mm} b^{\theta}_{\bar 2, \bar 3} = \sin r(\cos\phi, \sin\phi),
	\end{align}
	and
	\begin{equation}	
		G_{,r} = \sum^3_{L =1}\frac{-a_L\cos r}{(1-a_L^2\sin r)^{\frac{1}{2}}}A(r_L) = - A(r_{\bar 1}).
	\end{equation}
We have that
	\begin{align}
		G_{,\phi} &=0,\\
		G_{,\theta} &= S_{\bar 2}b_{\bar 2}^{\theta}A(r_{\bar 2}) + S_{\bar 3}b_{\bar 3}^{\theta}A(r_{\bar 3}),
	\end{align}
so that
	\begin{equation}
		\frac{1}{\sin r}G_{,\theta}=-\cos\phi A(r_{\bar 2})-\sin\phi A(r_{\bar 3})=0,
	\end{equation}
since $\cos r_L=a_L\sin r = 0$, so that $r_L=\frac{\pi}{2}$ ($L=\bar 2, \bar 3$), and $A\left(\frac{\pi}{2}\right)=0$.

Hence, on $\theta=0$, we obtain
	\begin{equation}
		G_{,\theta}=G_{,\phi}=0; ~~\Theta=\left[F_{,r}+G_{,r}\right]^2.
	\end{equation}
This implies that there is no influence from $BH_5 - BH_8$ on 	$\Theta$ on
$\theta = 0$. We also note that in this case the distance from $BH_1$ to $BH_2$ is $\pi$
(with half way point $\pi/2$).
That is, for $\theta=0$, and using $r_{\bar 1}-\frac{\pi}{2}-r$, we have that
	\begin{align} \label{E9}
		\sqrt{\Theta} &= F_{,r} + G_{,r} = A(r) - A(\frac{\pi}{2}-r)\\ \nonumber
&=-\frac{1}{\sin r} \left[\frac{-\cos r}{\sin^2 r}\big[\sinh K(\pi-r) + \sinh Kr\big] + K\big[\cosh K(\pi-r)-\cosh Kr\big]\right]\\ \nonumber
	&+\frac{1}{\cos r}\bigg\{\tan r\left[\sinh K\left(\frac{\pi}{2}+r\right)+\sinh K\left(\frac{\pi}{2}-r\right)\right]\\ \nonumber
	&+K\left[\cosh K\left(\frac{\pi}{2}+r\right) - \cosh K\left(\frac{\pi}{2}-r\right)\right]\bigg\}.
	\end{align}
We note that there is no remaining angular dependence in this expression.
We also remark that $r=r_{gh}=\frac{\pi}{4}$ is a solution of $\Theta=0$ for all $K$. We demonstrate in Figure 1 that this is the only root.

Note that $\cos r_{\bar 1}=a_{\bar 1}\sin r = \sin r$, so that for $r=\frac{\pi}{4}-x$, $r_{\bar 1} = \frac{\pi}{4}+x$ and we can do a small $x$ approximation close to $r\cong\frac{\pi}{4}$. We find that in this case
	\begin{equation}\label{doubledaggar}
		\Theta^{\frac{1}{2}}
		\cong -\frac{1}{\sqrt2} e^{\frac{3}{4}K\pi}\bigg\{2(3+2K+K^2)x - (7+8K+2K^2)x^2\bigg\}, 
	\end{equation}
\newpage
\begin{figure}[h!]
	\includegraphics[width=\linewidth]{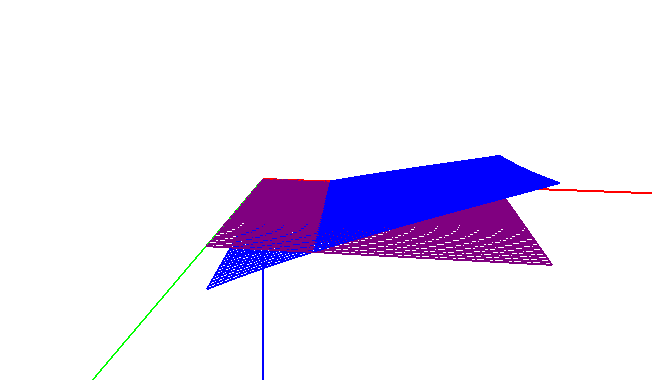}
	\caption{The roots of
$A(r) - A(\frac{\pi}{2}-r)$
in Eq.
(\ref{E9}).}
	\label{fig1}
\end{figure}

\noindent where various terms are included (even in the LKA approximations where various terms have similar exponential factors). Of course, $\Theta=0$ when $x=0$, which corresponds to $r_{gh}=\frac{\pi}{4}$, which is okay in the case $\theta=0$ when the distance from BH$_1$ to BH$_2$ is $\pi$. (But this implies that we should also study the case $\theta\neq0$.) If $x\neq 0$, we find that at $r_{gh}$,
	\begin{equation}
	x=\frac{2\big(3+2K+K^2\big)}{7+8K+2K^2}.
	\end{equation}
We note that as $K$ increases, the value for $x$ decreases (consistent with the earlier comments), but for $K$ in the physical range this gives rise to a nonsensical result (e.g. $r_{gh}$ is larger than the cell size), and one for which the approximation scheme breaks down. For modest values of $K$, Eq. \eqref{doubledaggar} receives corrections of the order of
	$$e^{-\frac{K\pi}{2}}\cdot2\big(3-2K+K^2\big)x,$$
which is always positive and leads to a small increase in the derived value of $r_{gh}$. We conclude that this analysis is inconsistent for $x\neq0$ (in particular, all approximations break down), and a value of $r_{gh}$ other than $r_{gh}=\frac{\pi}{4}$ is not possible. An attempt to analyze the case $r=\frac{\pi}{2}-x$ ($\frac{\pi}{2}$ is the half way point between BH$_1$ and BH$_2$) for small $x$ leads to inconsistencies.

\newpage 

\subsection{The case $\phi=\frac{\pi}{2}$} Since $\phi=\frac{\pi}{2}$ always in this case, we choose BH$_7$ ($\frac{\pi}{2},\frac{\pi}{2},\frac{\pi}{2}$) so that $a_{\bar 3}=a_7=-\sin\theta\sin\phi$ (and the distance from BH$_1$ to BH$_3$ and BH$_7$ is $\frac{\pi}{2}$). Recall that $A(\pi-r)=-A(r)$ and $A(\frac{\pi}{2})=0$. In this case, we thus have that:
\begin{table}[h!]
  \begin{center}
    \label{tab:table1}
    \begin{tabular}{r|c|c|c|c|c|c}
       & $a_{L}$ & $r_L$ & $\frac{b_L^r}{\cos r}$ & $\frac{b_L^{\theta}}{\sin r}$ & $\frac{b_L^{\phi}}{\sin r}$ & $S_L$\\
       \hspace{2mm} & \hspace{2mm} & \hspace{2mm} & \hspace{2mm} & \hspace{2mm} & \hspace{2mm} & \hspace{2mm}\\
      \hline
      $\bar 1$ & $\cos\theta$ & $r_{\bar 1}$ & $\cos\theta$ &  $-\sin\theta$ & $0$ & $S_{\bar 1}$ \\
      \hspace{2mm} & \hspace{2mm} & \hspace{2mm} & \hspace{2mm} & \hspace{2mm} & \hspace{2mm} & \hspace{2mm}\\
      $\bar 2$ & $0$ & $\frac{\pi}{2}$ & $0$ & $0$ & $-\sin\theta$ & $-1$ \\
      \hspace{2mm} & \hspace{2mm} & \hspace{2mm} & \hspace{2mm} & \hspace{2mm} & \hspace{2mm} & \hspace{2mm}\\
      $\bar 3$ & $-\sin\theta$ & $\bar r_3$ & $-\sin\theta$ & $-\cos\theta$ & $0$ & $S_{\bar3}$ \\
    \end{tabular}
  \end{center}
\end{table}\\
where $S_L=-\big(1-a_L^2\sin^2r\big)^{\frac{1}{2}}$, and $\cos r_L=a_L\sin r$, where $a_{\bar 3}=-\tan\theta a_{\bar 1}$. 

From $G_{,\chi}=\sum_{L=\bar 1}^{\bar 3}S_L b_L^{\chi} A(r_L)$, we then obtain
	\begin{align} \label{eqnG1}
		G_{,r}&=\frac{-\cos\theta\cos r}{\left(1-\cos^2\theta\sin^2 r\right)^\frac{1}{2}}\left[A(r_{\bar 1})+\frac{\left(1-\cos^2\theta\sin^2 r\right)^\frac{1}{2}}{\left(1-\sin^2\theta\sin^2 r\right)^\frac{1}{2}}\cdot (-\tan\theta) A\left(r_{\bar 3}\right)\right],\\ 
		G_{,\theta}&=\frac{\sin\theta\sin r}{\left(1-\cos^2\theta\sin^2 r\right)^\frac{1}{2}}\left[ A(r_{\bar 1})+\frac{\left(1-\cos^2\theta\sin^2 r\right)^\frac{1}{2}}{\left(1-\sin^2\theta\sin^2 r\right)^\frac{1}{2}}\cdot (-\cot\theta) A\left(r_{\bar 3}\right)\right], \label{eqnG11}
	\end{align}
and $G_{,\phi}=0$. Note that for $\theta=\frac{\pi}{4}$ ($r_{\bar 3}=\pi-r_{\bar 1}$) we have that
	\begin{align}
		G_{,r}&=\frac{-2\cos\theta\cos r}{\left(1-\cos^2\theta\sin^2 r\right)^\frac{1}{2}}\left[ A\left(r_{\bar 1}\right)\right],\\
		G_{,\theta}&=\frac{2\sin\theta\sin r}{\left(1-\cos^2\theta\sin^2 r\right)^\frac{1}{2}}\left[ A\left(r_{\bar 1}\right)\right].
	\end{align}
Now, $\cos r_{\bar 1}=\cos\theta\sin r$, $\cos r_{\bar 3}=-\sin\theta\sin r$. Writing $r=r_0+x$, from \eqref{dagger} we have that
	\begin{align}
	r_{\bar 1}&\cong R_{\bar 1} + P_{\bar 1}x; \qquad R_{\bar 1} = \arccos(\cos\theta\sin r_0), \qquad P_{\bar 1}=\frac{-\cos\theta\cos r_0}{\sqrt{1-\cos^2\theta\sin^2 r_0}},\\
	r_{\bar 3}&\cong R_{\bar 3} + P_{\bar 3}x;\qquad R_{\bar 3}=\arccos(-\sin\theta\sin r_0), \quad P_{\bar 3}=\frac{\sin\theta\cos r_0}{\sqrt{1-\sin^2\theta\sin^2 r_0}}.
	\end{align}

The solution for $\Theta=0$ when $\theta=0$ is $r_0=r_{gh}=\frac{\pi}{4}$ from earlier.
For small $\theta$, we write $\theta=\delta\theta$. We can neglect $G_{,\theta}$ and so $\sqrt{\Theta}\cong F_{,r} + G_{,r}$. We then find that
	\begin{equation}
		G_{,r}=-\left[ A\left(r_{\bar 1}\right)-\frac{\delta\theta}{\sqrt{2}} A\left(r_{\bar 3}\right)\right],
	\end{equation}
where $r_{\bar 1}=\frac{\pi}{4}-x$ and $r_{\bar 3}=\frac{\pi}{2} + \frac{1}{\sqrt2}\delta\theta$.

From Eq. \eqref{dagger}, we have that for $\alpha=\frac{\pi}{4}$ (so that $\bar r = \frac{\pi}{4} + \beta x$, where $\beta=1$ and $\beta=-1$ corresponds to $r$ and $r_{\bar 1}$, respectively):
\begin{align}
A\left(\frac{\pi}{4} + \beta x\right)= & A\left(\frac{\pi}{4}\right)  \\ \nonumber
		& +\beta x \left\{\sqrt{2}\left[3\left(\sinh \frac{3K\pi}{4} - \sinh \frac{K\pi}{4}\right) + K\left(\cosh \frac{K\pi}{4} - \cosh \frac{3K\pi}{4}\right)\right]\right. \\ \nonumber
		& \left.-\sqrt{2}K\left[\left(\cosh\frac{\pi K}{4} + \cosh\frac{3K\pi}{4}\right) + K\left(\sinh \frac{3K\pi}{4} + \sinh\frac{K\pi}{4}\right)\right]\right\}.
		\end{align}
And $r_{\bar 3}=\frac{\pi}{2} + \frac{1}{\sqrt2}\delta\theta$, so that from \eqref{dagger} with $\alpha = \frac{\pi}{2}$ and $\beta x \equiv \frac{1}{\sqrt2} \delta\theta$, we find that
	\begin{equation}	
		A\left(r_{\bar 3}\right) \cong \left\{\sqrt2(1+K)\sinh\frac{K\pi}{2}\right\}\delta\theta, \nonumber
	\end{equation}
since $A\left(\frac{\pi}{2}\right)=0$.
Now,
	\begin{align}
		\Theta^{\frac{1}{2}} = F_{,r} + G_{,r} &= A(r) - \left[A\left(r_{\bar 1}\right) - \frac{\delta\theta}{\sqrt2} A\left(r_{\bar 3}\right)\right] \nonumber \\
		&= 2\sqrt2\cdot I(K)x + (1+K)\sinh \frac{K\pi}{2}\left(\delta\theta\right)^2
	\end{align}
(as expected -- due to the choice of $r_0=\frac{\pi}{4}$ -- the $o(x^0)$ terms are zero), where
	\begin{align}
		I(K) \equiv \left(\sinh \frac{K\pi}{4} + \sinh \frac{3K\pi}{4}\right)\left(3+K^2\right) + 2K\left(\cosh \frac{3K\pi}{4} - \cosh \frac{K\pi}{4}\right).
	\end{align}

We note that ($\cosh \frac{3K\pi}{4} - \cosh \frac{K\pi}{4}$) is always positive (e.g., $\sim \frac{11}{4}$ for $K=\frac{2}{\pi}$ and $\sim 20$ for $K=\frac{4}{\pi}$) and thus $I(K)>0$. In the LKA,
	\begin{equation}
		I(K) \sim \frac{1}{2}\left(3+2K+K^2\right)e^{\frac{3K\pi}{4}},
	\end{equation}
which increases with $K$! Hence $\Theta^\frac{1}{2}=0$ implies that
	\begin{equation}
		x_{gh} = \frac{-(1+K)\sinh \frac{K\pi}{2} (\delta\theta)^2}{2\sqrt2 \cdot I(K)},
	\end{equation}
which is always negative. Therefore,  as $\delta\theta$ increases $x_{gh}$ is negative and decreases. (Note that for $K=\frac{2}{\pi}$, $x_{rg}\simeq - \frac{1}{20}(\delta\theta)^2$, and for $K=\frac{4}{\pi}$, $x_{gh} \simeq -\frac{1}{50}(\delta\theta)^2$). Therefore, {\em{around}} $\theta=0$:
	$$r_{gh}=\frac{\pi}{4}+x_{rg},$$
and the GH decreases below the critical value $\frac{\pi}{4}$. As $\delta\theta$ changes, $x_{gh}$ is negative and ($-x_{gh}$) increases.

\newpage

\subsection{The case $\theta=\frac{\pi}{2}$}

The distance from BH$_1$ to BH$_5$ and BH$_7$ (i.e., we are taking $\bar 2=5$ and $\bar 3=7$, respectively (and $\bar 1=3$)), is $\frac{\pi}{2}$. Using the definitions for $S_L$, $b_L^\chi$, and $\cos r_L=a_L\sin r$ (and the expression $G_{,\chi}$) from earlier, we have that
\begin{table}[h!]
  \begin{center}
    \label{tab:table1}
    \begin{tabular}{r|c|c|c|c|c|c|c}
       & $a_{L}$ & $\theta=\frac{\pi}{2}$ & $r_L$ & $s_L$ & $\frac{b_L^r}{\cos r}$ & $\frac{b_L^{\theta}}{\sin r}$ & $\frac{b_L^{\phi}}{\sin r}$\\
       \hspace{2mm} & \hspace{2mm} & \hspace{2mm} & \hspace{2mm} & \hspace{2mm} & \hspace{2mm} & \hspace{2mm} & \hspace{2mm}\\
      \hline
      $\bar 1$ & $\cos\theta$ & $0$ & $\frac{\pi}{2}$ &  $-1$ & $0$ & $-1$ & $0$ \\
      \hspace{2mm} & \hspace{2mm} & \hspace{2mm} & \hspace{2mm} & \hspace{2mm} & \hspace{2mm} & \hspace{2mm} & \hspace{2mm}\\
      $\bar 2$ & $-\sin\theta\cos\phi$ & $-\cos\phi$ & $r_{\bar 2}$ & $S_{\bar 2}$ & $-\cos\phi$ & $0$ & $\sin\phi$ \\
      \hspace{2mm} & \hspace{2mm} & \hspace{2mm} & \hspace{2mm} & \hspace{2mm} & \hspace{2mm} & \hspace{2mm} & \hspace{2mm}\\
      $\bar 3$ & $-\sin\theta\sin\phi$ & $-\sin\phi$ & $r_{\bar 3}$ & $S_{\bar 3}$ & $-\sin\phi$ & $0$ & $-\sin\phi$ \\
    \end{tabular}
  \end{center}
\end{table}\\
Noting that $A\left(\frac{\pi}{2}\right)=0$ and assuming $0<\phi<\frac{\pi}{2}$, we obtain $G_{,\theta}=0$ and
	\begin{align}
		G_{,r} &= \frac{\cos r\cos\phi}{\sqrt{1-\cos^2\phi\sin^2 r}}\left(A\left(\bar r_2\right) + \frac{\left(1-\cos^2\phi\sin^2 r\right)^{\frac{1}{2}}}{\left(1-\sin^2\phi\sin^2 r\right)^{\frac{1}{2}}} \tan\phi \cdot A\left(r_{\bar 3}\right)\right),\\
		G_{,\phi} &= \frac{-\sin r\sin\phi}{\sqrt{1-\cos^2\phi\sin^2 r}}\left(A\left(\bar r_2\right)-\frac{\left(1-\cos^2\phi\sin^2 r\right)^{\frac{1}{2}}}{\left(1-\sin^2\phi\sin^2 r\right)^{\frac{1}{2}}} A\left(r_{\bar 3}\right)\right).
	\end{align}
If $\phi=\frac{\pi}{4}$ (as well as $\theta=\frac{\pi}{2}$), then $r_{\bar 3}=r_{\bar 2}$ where $\cos r_{\bar 2} = -\frac{1}{\sqrt 2}\sin r$. We then obtain
	\begin{align} \label{eqnG2}
		G_{,r} &= \frac{\frac{1}{\sqrt 2}\cos r}{\sqrt{1-\frac{1}{2}\sin^2 r}}\bigg[ A\left(\bar r_2\right) + A\left(r_{\bar 3}\right)\bigg] = \frac{\sqrt{2}\cos r}{\sqrt{1-\frac{1}{2}\sin^2 r}}A\left(r_{\bar 2}\right),\\
		G_{,\phi} &= \frac{-\frac{1}{\sqrt{2}}\cos r}{\sqrt{1-\frac{1}{2}\sin^2 r}}\bigg[ A\left(\bar r_2\right) - A\left(r_{\bar 3}\right)\bigg] = 0.
	\end{align}
Hence
	\begin{equation}
		\sqrt{\Theta}(K,r) \equiv F_{,r} + G_{,r}. \label{eqnroot}
	\end{equation}
The roots of this Eq. (for which $r_{gh}$	is always less than $\pi/4$) are displayed in Figure 2.
	Doing an expansion for $\phi$ around $\frac{\pi}{4}$ (i.e., $\phi=\frac{\pi}{4}+\delta\phi$), and neglecting $G_{,\phi}$ (which is very small), we find that $r_{\bar 3}-r_{\bar 2} \sim o(\delta\phi)$ and $\Theta^{\frac{1}{2}} \cong \bar{A}(r,K) + \beta(r,K)\delta\phi$, where $\beta\neq 0$ and the linear correction is non-zero (and of either sign) for general $r$ and $K$.
\newpage
\begin{figure}[h]
	\includegraphics[width=\linewidth]{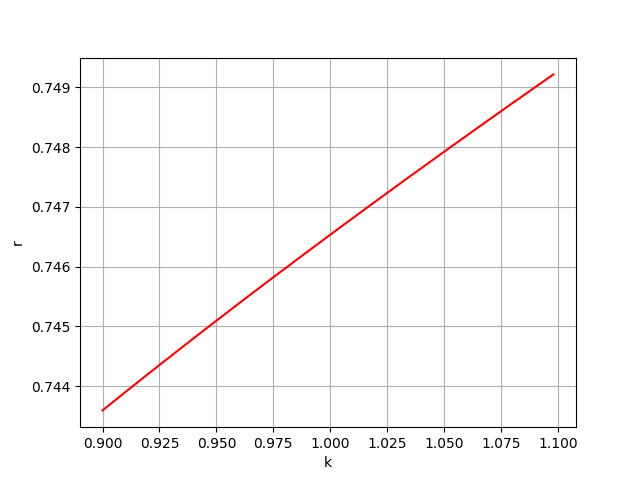}
	\caption{The roots of \eqref{eqnroot} for $F_{,r} + G_{,r}$.}
	\label{fig2}
\end{figure}

\newpage

\section{Conclusions}

In order to study whether black holes can persist in a Universe that undergoes a cosmological bounce, we have investigated what happens to the number density of black holes as the minimum of expansion is approached and whether it is
possible that the filling factor, $F$, never reaches $1$. 
To this end we have considered the class of
black hole lattice models in a hyperspherical cosmology which  
undergo a dynamical  bounce due to a scalar field \cite{bounce}.
We have derived time evolving solutions 
of instantaneously-static models
perturbatively. And we have utilized the notion of a GH, which can be simply determined by curvature invariants, to characterize the black hole horizons.

We have focused on the particular case of
eight regularly-spaced black holes with equal masses and considered the more relevant case of $\kappa > 1$. Indeed, we have explicitly
computed the invariants for these exact solutions  necessary to determine the conditions for the existence of a GH.

\subsubsection{Summary of $\kappa > 1$ case}

We have derived analytical expressions for $\Theta$, where the root of $\Theta=0$ defines the GH, $r_{gh}$. We are most interested in the  physical values $0.45 < K \lesssim  1.09$ (in which the black holes were found not to merge in the analysis of CCC), but we also considered qualitative results in the small r analysis
and for larger values of $K$ in the LKA.

We considered the specific case $\theta=0$ in detail. We obtained the exact expression for $\Theta=0$ in Eq. \eqref{E9}, which has the exact solution $r_{gh}=\frac{\pi}{4}$ for all values of $K$, which is much less than half the distance $\left(\frac{\pi}{2}\right)$ between BH$_1$ and BH$_2$ on $\theta=0$.
We then studied the case $\phi=\frac{\pi}{2}$ (where half the distance from BH$_1$ to BH$_3$ and BH$_7$ is $\frac{\pi}{4}$). In this case $G_{,\phi}=0$ and $G_{,\theta}$ and $G_{,r}$ are given by Eqs. \eqref{eqnG1} and \eqref{eqnG11}. We then considered the subcase of $\theta=0+\delta\theta$, and found that the GH always decreases from its value of $\frac{\pi}{4}$ at $\theta=0$ as $|\delta\theta|$ increases.
Finally, we considered the case $\theta=\frac{\pi}{2}$ ($G_{,\theta}=0$, and $G_{,r}$ and $G_{,\phi}$ given by Eqs. \eqref{eqnG2} and \eqref{eqnG22}), and briefly discussed the subcase $\phi=\frac{\pi}{4} + \delta\phi$.

We conclude that the black holes do not merge before nor at the bounce in these models

\subsubsection{Discussion}

If we assume that the area of a black hole horizon in a contracting cosmology
is increasing,
then during contraction the distance between the black holes, $r_{sep}$ (which is $\pi/2 $ at the bounce at  $t=0 $ in our models), is decreasing and hence if F at $t=0$ is less than unity, then necessarily  $F<1 $ before the bounce.
In addition, the Schwarzchild mass of a black hole  (and hence the black hole horizon)
increases as $\kappa$ increases, and so there is an upper limit on $\kappa$ in order
for $F<1$ at the beginning of the collapse in the calculations; i.e., $\kappa$ cannot be
arbitrarily large.

For an equally spaced BHL with equal separation, $r_{sep}$, suppose that $r_{crit}$ is the separation value at the bounce such that all of the black holes merge at
or before the bounce. That is, all black holes will merge if   $r_{sep} < r_{crit} $, and none will merge if  $r_{sep} > r_{crit}$. If we then have a distribution of black holes with an arbitrary (random) spacing  
 $r_{sep} $, there will be hierarchical merging depending the relative distances between the black holes. Let the average distance between the black holes in this distribution be $r_{av}$. Suppose that $r_{av} < r_{crit}$, then it is reasonable that black holes with  $r_{sep}$ less than or approximately equal to $r_{av} < r_{crit} $  will merge and that black holes for  $r_{sep} $ greater than or approximately equal to $r_{av} > r_{crit} $ will not merge. That is,  $r_{crit} $ will be a characteristic scale such that above which (on average in some sense) the black holes in the random distribution will persist.
Therefore, it might be expected that in general the results 
for randomly distributed black holes would be qualitatively the same as for 
the case of equally spaced BHL discussed above; i.e., for  $r_{sep} > r_{av}> r_{crit} $, there are
black holes that will not merge before the bounce.

\newpage

\section{Appendix: horizons for vacuum black holes}

Following \cite{DurkClif}, the vacuum constraint Eqs. are:
\begin{equation}
      \mathcal{R} + K^2 - K_{ij}K^{ij}  = 0, ~~~
     D_j(K_i^{\,\,j} - \delta_i^{\,\,j}K)  = 0 \, ,
\end{equation}
where
\begin{equation}
\label{eq:ext}
 K_{ij} \equiv -\frac{1}{2}\partial_t h_{ij} \, ,
\end{equation}
and $\partial /\partial t$ is a vector orthogonal to the initial hyper-surface, which is choosen to be symmetric relative to a reversal of time.
This means that  this surface is static at this instant and that the extrinsic curvature consequently vanishes and that the second constraint above is automatically satisfied and we have that $\mathcal{R} = 0$.  If we now rescale the metric conformally (i.e., $h_{ij}=\Omega^4 \tilde{h}_{ij}$), then
\begin{equation}
\mathcal{R} = \Omega^{-4}\tilde{\mathcal{R}} - 8 \Omega^{-5} \tilde{D}^2 \Omega = 0 \, \label{eqnG22},
\end{equation}
where $\tilde{\mathcal{R}}$ is the Ricci  curvature scalar of the conformal 3-space and $\tilde{D}_i$ denotes covariant differentiation on it. If we now choose $\tilde{h}_{ij}$ so that $
d\tilde{s}^2 = \Omega^4[d\xi^2 + \sin^2\xi ( d\theta^2 + \sin^2\theta d\phi^2)]$,
then   $\tilde{\mathcal{R}}=6$ and both of the Eqs. above are satisfied at the moment of time-reversal, and $\Omega$ satisfies
\begin{equation}
\label{eq:linear}
\tilde{D}^2 \Omega = \frac{3}{4} \Omega  \, .
\end{equation}
This equation is linear in $\Omega$, and has solutions in the form of terms similar to $\Omega \propto 1/\sin(\xi/2)$, where each such term corresponds to an additional black hole positioned at a different location on the conformal 3-sphere; that is,
additional terms in the conformal factor $\Omega$ can be included by a rotation of the 3-sphere by an arbitrary angle and by then adding a new term of the form $1/\sin(\xi/2)$ in the resulting coordinates, thereby generating by summation the conformal factor $\Omega$ representing multiple black holes:
\begin{equation} \label{psi0}
\Omega(\xi, \theta, \phi) = \sum_{i=1}^{N} \frac{\sqrt{\tilde{m}_i}}{2f_i(\xi, \theta, \phi)} \, ,
\end{equation}
where $N$ represents the total number of black holes, the $\tilde{m}_i$ are arbitrary constants, and
\begin{equation}
f_i = \sin\left({\frac{1}{2}\arccos(h_i)}\right) \, ,
\end{equation}
where the $h_i$ are defined by
\begin{equation}\label{eq:functions}
h_i = w_i \cos{\xi} +  x_i \sin{\xi}\cos{\theta} + y_i \sin{\xi}\sin{\theta}\cos{\phi} + z_i  \sin{\xi}\sin{\theta}\sin{\phi} \, ,
\end{equation}
for a divergent term  at  $(w_i, x_i, y_i, z_i)$ (which represents the position in the initial data of a point-like mass).

We can consequently construct the (exact) initial data for a bouncing cosmology containing $N$ distinct black holes \cite{DurkClif}. Assuming that these black holes are the same distance from each of their closest neighbours, 
they form a perfect regular lattice. 
The conformal 3-sphere could then be tiled with 
(one of six possible) regular polyhedra by positioning a Schwarzschild (black hole) mass at the centre of each \cite{clifton2}. 
These models can be generalised by relaxing the assumption of a perfect equidistant spacing.

Let us review the 
identification of the locations of the apparent black hole horizons in the 
cosmologies \cite{DurkClif}
(which are used to determine when the horizons of nearby black holes remain distinct). We note that in the investigation of the clustering of two adjacent black holes, as the objects approach each other an additional apparent horizon may appear which subsumes them both; when this happens the individual black holes retain their own horizons,  but also acquire a new shared horizon.

\subsection{Locating apparent horizons}

Apparent horizons are  2-dimensional MOTS, defined mathematically by the vanishing of the expansion of the (outwardly directed) null normal to the surface; i.e., $\nabla_\mu k^{\mu} =0$. The position of apparent horizons can be obtained by the following two different methods.

\paragraph{The Area Method:} In the case of time symmetry, for all MOTS the (outward-pointing) light-like normal vector can be decomposed in terms of a  time-like vector $u^{\mu}$ and a space-like vector $e_1^{\mu}$ (orthogonal to $u^{\mu}$), such that
\begin{equation} \label{exhor}
\nabla_{\mu} k^{\mu}  = \frac{1}{\sqrt{2}} \nabla_{\mu} u^{\mu}+\frac{1}{\sqrt{2}} \nabla_{\mu} e_1^{\mu} = 0 \, .
\end{equation}
The vanishing of  the extrinsic curvature of the initial data
then implies that $\nabla_{\mu} u^{\mu}=0$ and  $\nabla_{\mu} e_1^{\mu} =0$, so that $e_1^{\mu}$ is orthogonal to that apparent horizon, which is
consequently an extremal (minimal) closed surface in the 3-space. 
If the gravitational field near each black hole is approximately spherically symmetric, 
then the position of the apparent horizon can be estimated by determining the value of $\xi$ that minimizes 
\begin{equation}
\label{eq:area}
A(\xi) = \int_0^{2\pi} \int_0^{\pi} \Omega^4 \sin^2(\xi) \sin(\theta) d\theta d\phi \, .
\end{equation}
This method is straightforward from a computational point of view. However, it is only reliable when the black hole horizon is approximately spherically symmetric (which is not the case when the black holes are close together).

\paragraph{The Weyl Tensor Method:} In the case of  extrinsically flat initial data, the Ricci identities and the Gauss Eq. can be used to obtain \cite{DurkClif,clifton2}:
\begin{equation} \label{ER}
E_{\mu\nu} =  \mathcal{R}_{\mu\nu} \, ,
\end{equation}
where $E_{\mu \nu}$ is the electric part of the Weyl tensor relative to the 4-velocity $u^{\mu}$, and $\mathcal{R}_{\mu\nu}$ is the 3-dimensional Ricci tensor.  $\mathcal{R}_{\mu\nu}$ can then be calculated explicitly, and utilizing the Bianchi identities and  Eq. (\ref{exhor}), the location of the apparent horizon
can now be the identified by the simple necessary condition $\mathbf{e}_1(E^{11}) =0$ along locally rotationally symmetric curves \cite{DurkClif} 
(where  we choose coordinates so that the frame derivative $\mathbf{e}_1 = \Omega^{-2} \partial_{\xi}$ points along these curves and  $ E^{11}$
is a frame component of the electric Weyl tensor)
or, equivalently,
\begin{equation} \label{eq:weyl}
\frac{1}{\Omega^2}\frac{\partial}{\partial \xi} (\Omega^{-4} {\mathcal R}_{\xi \xi}) = 0 \, ;
\end{equation}
i.e., the MOTS are found at points where $\Omega^{-4} {\mathcal R}_{\xi \xi}$ is optimized.

The area method gives a reliable estimate for the position of the shared apparent horizon, at least for small parameter values \cite{DurkClif,clifton2}.
The Weyl tensor method (when the curves that are rotationally symmetric locally intersect the horizon at points parameterised by $\xi$)
is expected to be more accurate than the area method, particularly when the horizon is not spherical. In the applications in this paper the 
Weyl tensor method appears to be related to identifying the GH, and it could be speculated that this might be the case in a wider context.

\newpage

\paragraph{Acknowledgments:} 
I would like to thank T. Clifton and B. Carr for helpful discussions, and Nick Layden for help with the figures. Financial support from NSERC of Canada is gratefully acknowledged.

%% \cite{preprint}

\end{document}